\documentclass[]{emulateapj}

\shorttitle{The Shapes of Galaxy Groups}
\shortauthors{Robotham et al.}

\begin{document}


\title{The Shapes of Galaxy Groups - Footballs or Frisbees ?}


\author{Aaron Robotham, Steven Phillipps}
\affil{H.~H.~Wills Physics Laboratory, University of Bristol,\\
       Tyndall Avenue, Bristol, BS8 1TL, United Kingdom}
\email{A.Robotham@bristol.ac.uk}

\and

\author{Roberto De Propris}
\affil{Cerro Tololo Inter-American Observatory, Casilla 603, La Serena, Chile}



\begin{abstract}

We derive probability density functions for the projected axial ratios of the real and mock 2PIGG galaxy groups, and use this data to investigate the intrinsic three dimensional shape of the dark matter ellipsoids that they trace. As well as analysing the raw data for groups of varying multiplicities, a convolution corrected form of the data is also considered which weights the probability density function according to the results of multiple Monte-Carlo realizations of discrete samples from the input spatial distributions. The important effect observed is that the best fit distribution for all the raw data is a prolate ellipsoid with a Gaussian distribution of axial ratios with $\bar{\beta}=0.36$ and $\sigma=0.14$, whilst for the convolved data the best fit solution is that of an oblate ellipsoid $\bar{\beta}=0.22$ and $\sigma=0.1$. Previously only prolate distributions were thought compatible with the data, this being interprated as evidence of filamentary collapse at nodes. We also find that even after allowing for the sampling effects, the corrected data is better fit using separate multiplicity bins, which display a trend towards more spherical halos in higher multiplicity groups. Finally, we find that all results in the real data are in good agreement with the mock data from $\Lambda$CDM simulations, KS tests showing that all comparative data have been drawn from the same distributions within the $1\sigma$ confidence limits.

\end{abstract}


\keywords{galaxies: clusters: general; galaxies: halos}


\section{Introduction}

The spatial distribution of galaxies traces the shape of the dark matter
potential in which they are embedded. Simulations show that dark matter
halos are not spherical, as one would naively expect because of dark matter's
non-dissipational nature, but are strongly flattened triaxial ellipsoids
\citep{dubinski91}. Although prolate and oblate shapes are equally likely in the
simulations, dissipative infall of baryonic gas eventually forces the halo
shapes toward pronounced oblateness with axial ratios $b/a > 0.7$ \citep{katz91,
dubinski94,combes02}.

Because dark matter structures evolve self-similarly it is possible to derive
the general shape of dark halos from analysis of single test objects. The
first such attempts were carried out on rich clusters, whose shapes were found
to be triaxial and strongly prolate \citep{plionis01}: however the non-linear
evolution of these objects may play a role in shaping their density distributions
\citep[e.g][]{binney79}. Indeed there is evidence that the more high-density systems
are more spherical than low-density objects \citep[random gaussian field work of][]{bardeen86}. The observed trends are the opposite of what is observed in simulations \citep[e.g.][]{kasun05,allgood06}.

The distribution of Milky Way satellites shows a strongly oblate and flattened
distribution \citep{ruzicka07}, which is also consistent with analysis of the
orbital planes of the Sagittarius dwarf \citep{johnston05} and the Monoceros
stream \citep{penarrubia05}. However, star counts \citep{lemon04} and analysis
of the stellar stream of the Sagittarius dwarf \citep{ibata01} support a 
spherical halo. It is also uncertain whether the dwarf satellites can be
used as test particles, as they may not originate from cosmological sub-structures
\citep{kroupa05}.

Groups of galaxies are likely to be the best testbeds to study the shapes of
dark matter halos. Large group catalogs are now available from redshift surveys
(e.g., the 2dF Percolation Inferred Galaxy Groups [2PIGG] of \citealt{eke04} and
the group catalog of \citealt{merchan05} from Data Release 3 of the Sloan Digital
Sky Survey) and these allow a statistical approach to the study of the shapes of
groups and the shape dependence on richness, multiplicity and dynamical evolution.
Early studies marginally favoured prolate shapes \citep{fasano93,orlov01}, but not at the exclusion of oblate solutions. The most recent study of 2PIGG groups by \cite{plionis06} favours prolate groups, but is obtained by means of a multiplicity cut and as such represents a different method to the one presented here. Prolate results indicate that the
original (oblate ?) shapes may have been strongly modified during gravitational 
collapse, or that filamentary collapse is in fact being witnessed.

In this paper we carry out a detailed analysis of the shape of 2PIGG groups,
coupled with extensive Monte Carlo simulations and comparison with groups 
extracted from the 2dF mock catalogs used by \cite{eke04} to optimize group
selection. In section 2 we present our method for analysing the shape of observed 
groups and in section 3 apply this to the real and mock 2PIGG catalog \citep{eke04}. Comparisons are made between the raw data and data that is corrected for the finite (indeed, often sparse) sampling by factors which we determine from Monte-Carlo simulations of suitably populated groups. Furthermore, the mock and real data are considered separately and KS tests are used to confirm their distributional similarities.

\section{Modelling Shape Distributions}

\begin{figure*}[t]
\plotone{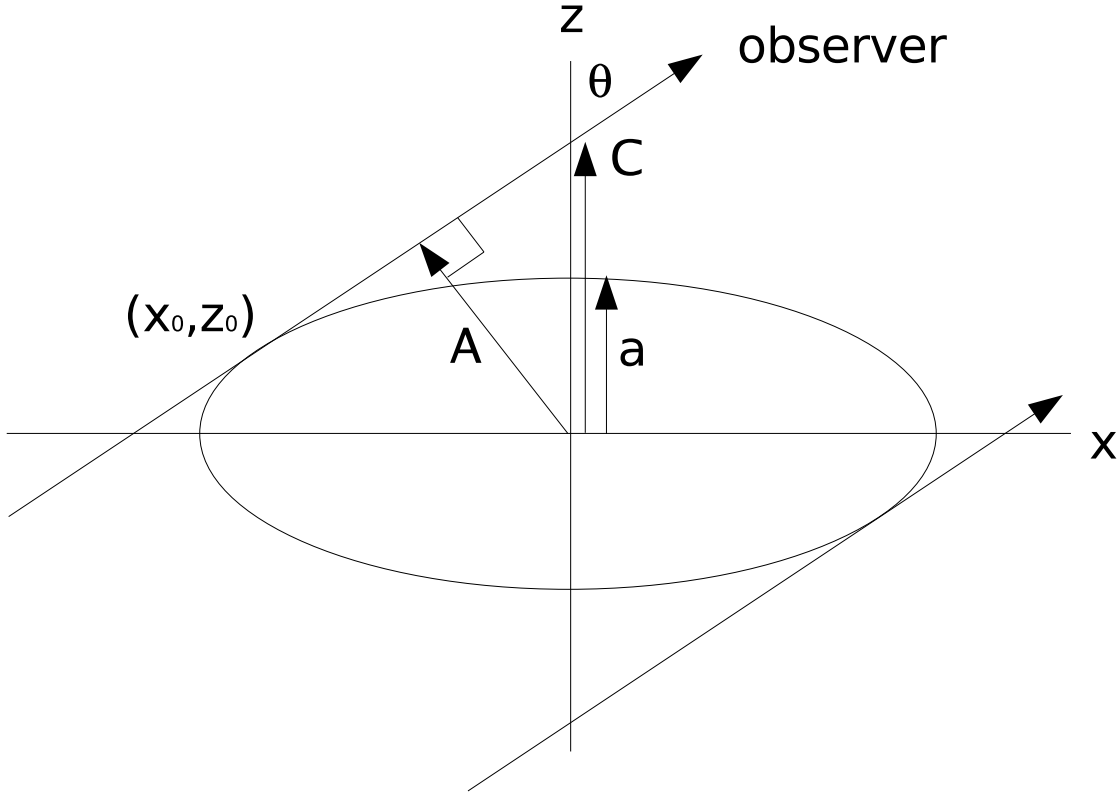}
\caption{Geometric description of ellipsoid viewing problem.}
\label{ellipse}
\end{figure*}

To extract useful information from our projected group shapes we first make some simplifying assumptions. With some theoretical justification \citep{jing02} we 
assume that our groups have three orthogonal axis with two of equal length, but 
allow our groups to be either prolate or oblate. This greatly simplifies the 
modelling compared to the case of general triaxial shapes. The problem is generally
one of inversion, for which there is no unique solution in case of triaxial shapes. However, it is still possible to recover whether the distribution is more prolate or oblate-like, as will be discussed later.

For now, we shall also make the assumption that we are able to measure the apparent group axial ratios perfectly. Obviously this is not the case in practice since the group projection is represented by discrete galaxies, but we consider the effect of this in detail later. Figure \ref{ellipse} describes the simplified problem we are considering. In the diagram shown the object is oblate and symmetric about the z axis. C is where the tangent to the observer (at the point $x_0$, $z_0$ in the x,z plane) cuts the z-axis, A is the distance between the origin and the tangent to the observer along the line normal to the tangent, {\it a} is where the ellipsoid cuts the z-axis and $\theta$ is the angle between the z-axis and the tangent to the observer. The axial ratio for this projected image is found by considering the distance between the tangent lines to the observer (2A) and the larger axis that is projected normal to this on the sky (twice the radius of the ellipsoid in the x,y plane). The ratio between these values will tell us the observed axial ratio for each given $\theta$ and ellipsoid shape.

Knowing this we can investigate what the probability density function (PDF) of axial ratios would be for any given ellipsoid evenly sampled in spherical coordinate parameter space. For a prolate or oblate object we have the standard form

\begin{equation}
(ux)^2+(uy)^2+z^2=a^2
\end{equation}

where $u>1$ for a prolate ellipsoid, $u<1$ for an oblate ellipsoid, and {\it a} would be the radius of the sphere for the case $u=1$.

We can then calculate the projected shape of the ellipsoid on the plane of the sky when viewed at an arbitrary angle $\theta$ by generalizing the results on the projected ellipticities of disc or elliptical galaxies from, e.g., \citet{hubble26}, \citet{sandage70} and \citet{binney81}; see appendix for details. The projected axis length A (Fig. \ref{ellipse}) is given by

\begin{equation}
A^2=a^2\left(\frac{a^2}{z_0^2}\right)\sin^2\theta=a^2\left(\frac{\cos^2\theta}{u^2}+\sin^2\theta\right)
\end{equation}

\noindent so the apparent axial ratio q (defined to be less than 1 for both oblate and prolate shapes: i.e. $q=\frac{uA}{a}$ or $\frac{a}{uA}$ respectively) is given by

\begin{eqnarray*}
u^2\sin^2\theta+\cos^2\theta&=&q^2 (oblate)\\ &=&\frac{1}{q^2} (prolate)
\label{trigdef}
\end{eqnarray*}

If we have a distribution of intrinsic shapes given by $n(u)$ and random orientations then we can show (again see appendix for details) that the distribution of apparent axial ratios will be

\begin{equation}
f(q)=q\int^q_0\frac{n(u)du}{[(1-u^2)(q^2-u^2)]^{0.5}} (oblate)
\end{equation}

\begin{equation}
f(q)=\frac{1}{q^2}\int^q_0\frac{\frac{1}{u^2} n(u)du}{[(1-\frac{1}{u^2})(q^2-\frac{1}{u^2})]^{0.5}} (prolate)
\end{equation}

\noindent since only values of $u<q$ can contribute. If we write the intrinsic axial ratio as $\beta<1$ (so $\beta=u$ for oblate and $\beta=1/u$ for prolate shapes), then for a given $\beta$, the number of systems with an observed $q$ between, say, $q_1$ and $q_2$ for oblate and prolate ellipsoids respectively is

\begin{eqnarray*}
N(\beta,q_1,q_2)=\frac{1}{\sqrt{1-\beta^2}}\int^{q_2}_{q_1}\frac{q dq}{\sqrt{q^2-\beta^2}}&=&\frac{\left[\sqrt{q_2^2-\beta^2}-\sqrt{q_1^2-\beta^2}\right]^{q_2}_{q_1}}{\sqrt{1-\beta^2}}\\
=\frac{1}{\sqrt{1-\beta^2}}\int^{q_2}_{q_1}\frac{\beta^2 dq}{q^2\sqrt{q^2-\beta^2}}&=&\frac{\left[\frac{\sqrt{q_2^2-\beta^2}}{q_2}-\frac{\sqrt{q_1^2-\beta^2}}{q_1}\right]^{q_2}_{q_1}}{\sqrt{1-\beta^2}}\\
\end{eqnarray*}

So for a distribution of true axial ratios $\tilde{n}(\beta)$, the apparent axial ratios for oblate ellipsoids follow

\begin{equation}
N(q_1,q_2)=\int^{q_2}_0\tilde{n}(\beta)\frac{1}{\sqrt{1-\beta^2}}\left[\sqrt{(q_2^2-\beta^2)}-\sqrt{(q_1^2-\beta^2)}\right]d \beta
\end{equation}

and for prolate ellipsoids

\begin{equation}
N(q_1,q_2)=\int^{q_2}_0\tilde{n}(\beta)\frac{1}{\sqrt{1-\beta^2}}\left[\frac{\sqrt{(q_2^2-\beta^2)}}{q_2}-\frac{\sqrt{(q_1^2-\beta^2)}}{q_1}\right]d \beta
\end{equation}

The simplest distribution to try would obviously be a delta function $\tilde{n}(\beta)=\delta(\beta_1)$ for some $\beta_1$, however a more sensible assumption for the underlying axial ratio function (or at least the first that should be tried) would be a Gaussian distribution, where

\begin{equation}
\tilde{n}(\beta)=\sqrt{\frac{2}{\pi\sigma^2}}exp {\left(-\frac{(\bar{\beta}-\beta)^2}{2\sigma^2}\right)}
\end{equation}

We will use this model in what follows.

\section{Methodology}

The largest publicly available group samples are the 2dF  Percolation-Inferred Galaxy Groups (2PIGG) catalog \citep{eke04}, selected from the 2dF Galaxy Redshift Survey (2dF-GRS -- \citealt{colless01}), the Yang 2dF group catalog \citep{yang05a}, and the Yang SDSS group catalog \citep{weinmann05}. For the purposes of this paper only the 2PIGG catalog and its associated mock catalog will be used; the other catalogs will be used in a more comprehensive paper (in preparation) comparing the different grouping methods. As in our previous work \citep{robotham06} we begin by selecting all  groups from this catalog with $0.05 < z < 0.10$. The $z=0.05$ lower limit is motivated by the small volume sampled by 2PIGG at low redshift (so that groups may not be representative), while at $z=0.10$ the 2dF GRS apparent magnitude limit begins to exclude even moderate luminosity galaxies ($M_B < -18$) from the sample (making the members less representative).

It is not trivial to decide how many discrete objects you require in an elliptically distributed sample to accurately determine the underlying shape. To  resolve this problem, Monte-Carlo tests were performed using a range of initial spatial distributions randomly populated by a varying number of discrete points. Using the moments method of \citet{carter80} to determine ellipticity (found to be preferential to the flattening technique of \citep{rood79} in \citep{plionis04}), these simulated discrete points were analysed and compared to the true elliptical distributions that produced them. This method of simulating and measuring was carried out 1,000 times for each multiplicity of group distributed with an ellipticity defined by $\varepsilon=1-\beta$ (i.e. an ellipticity of 0 would be a circle). The multiplicity was varied in steps of 5 up to 300, and $\varepsilon$ in steps of 0.05 from 0 up to 0.95. With this extensive sample available, comparisons of standard deviations in determined $\varepsilon_o$ for different simulated multiplicities were made versus the intrinsic $\varepsilon_i$ used to generate the samples. It was discovered that the observed ellipticity correlates almost linearly with the true ellipticity, error distributions are very close to Gaussian, and the standard error for a measured ellipticity is directly proportional to the axial ratio (when defined to be $<1$). As we find a near linear correlation between the observed and the true ellipticity, by fitting to the distribution of the fractional differences between the intrinsic and observed ellipticities as a function of the multiplicity ($n$) we can correct the ellipticity. The empirical correction function is

\begin{equation}
\varepsilon_i=1+\left(\frac{\varepsilon_o -1}{0.99294-\left(\frac{0.224438}{N}\right)-\left(\frac{0.11011}{N^2}\right)+1.22467\times10^{-5}N}\right)
\end{equation}

\begin{figure}[t]
\epsscale{0.80}
\plotone{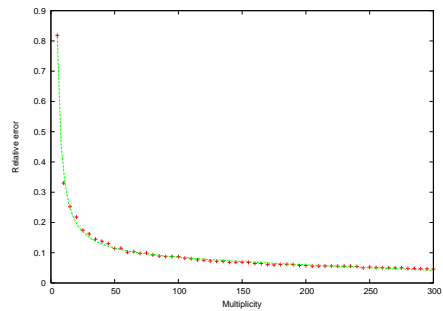}
\caption{Relative error of measured axial ratio, overplot with the best fit function.}
\label{relerr}
\end{figure}
 
However, this systematic correction is not required usually since it is so small compared to the standard error of the distribution of $\varepsilon_i$ for a given $\varepsilon_o$. Using the same data, a fit has been made of the relative axial ratio error against multiplicity (see figure \ref{relerr}). The standard error is given by

\begin{equation}
\sigma_N=\beta\left(0.0792+\left(\frac{2.0119}{N}\right)+\left(\frac{8.3096}{N^2}\right)-1.47\times10^{-4}N\right)
\end{equation}

This function is plotted against the data in figure \ref{relerr}. From this we see that independent of the underlying $\varepsilon_i$, 20 objects are required to have a relative axial ratio error of $20\%$. If we have 20 objects in a group with measured ellipticity of 0.2 ($\beta=0.8$), $\sigma$ will be 0.16, whilst for a more elliptical group with $\varepsilon_0=0.6$ ($\beta=0.4$), the error will be 0.08. There is obviously an issue with the observed distribution of ellipticities as we move towards 0; a simple but accurate model is that the distribution remains Gaussian into negative values for the ellipticity, but these probabilities are mirrored back across the origin. Accordingly, the distribution about a true ellipticity of 0 will appear to be a Gaussian cut in half, and the distribution about a true ellipticity of 0.4 is almost a pure Gaussian with $\sigma$ as stated. 

\begin{figure}[t]
\plotone{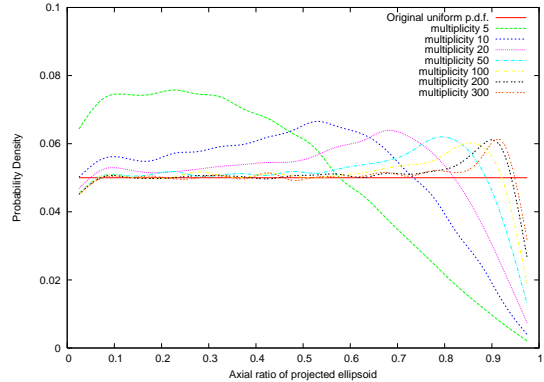}
\caption{Observed PDF for different multiplicities plotted together with the uniform input PDF used for the Monte-Carlo tests.}
\label{pdfax}
\end{figure}

\begin{figure*}[t]
\plottwo{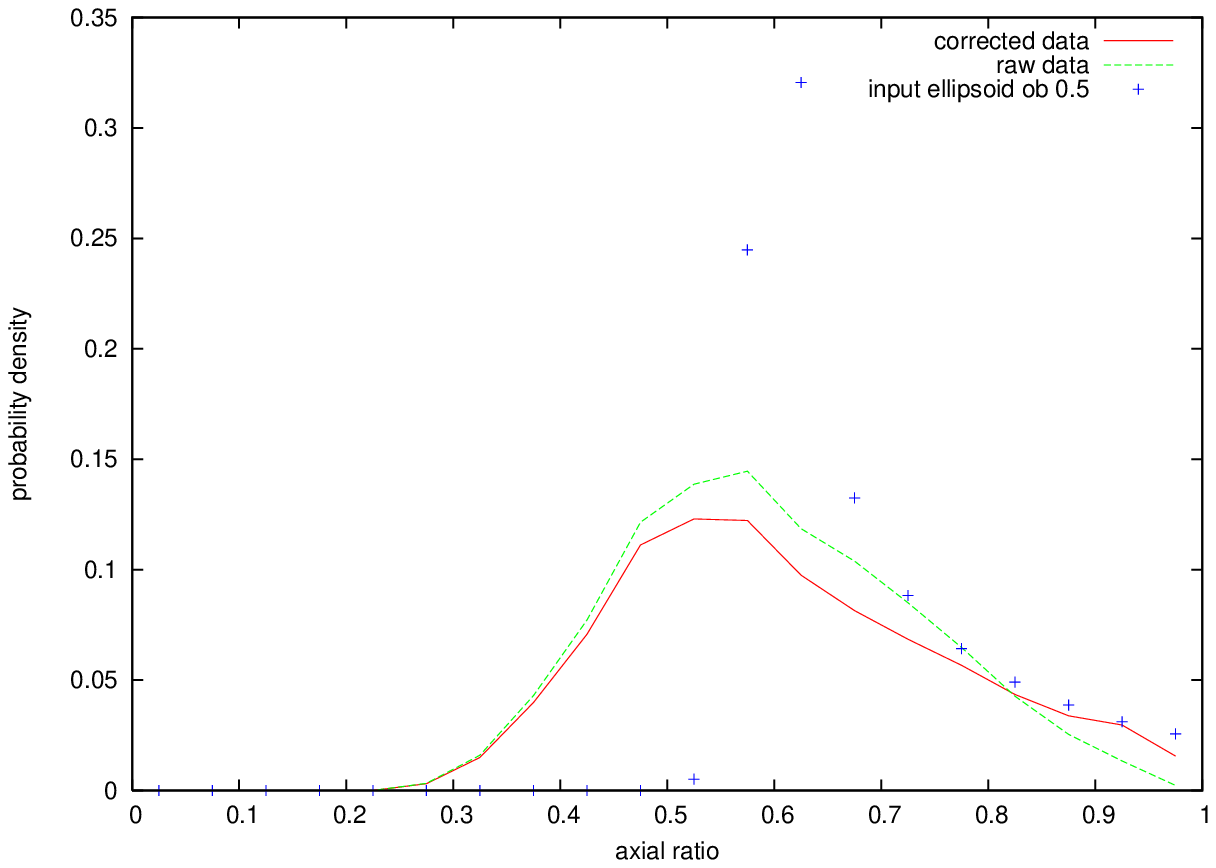}{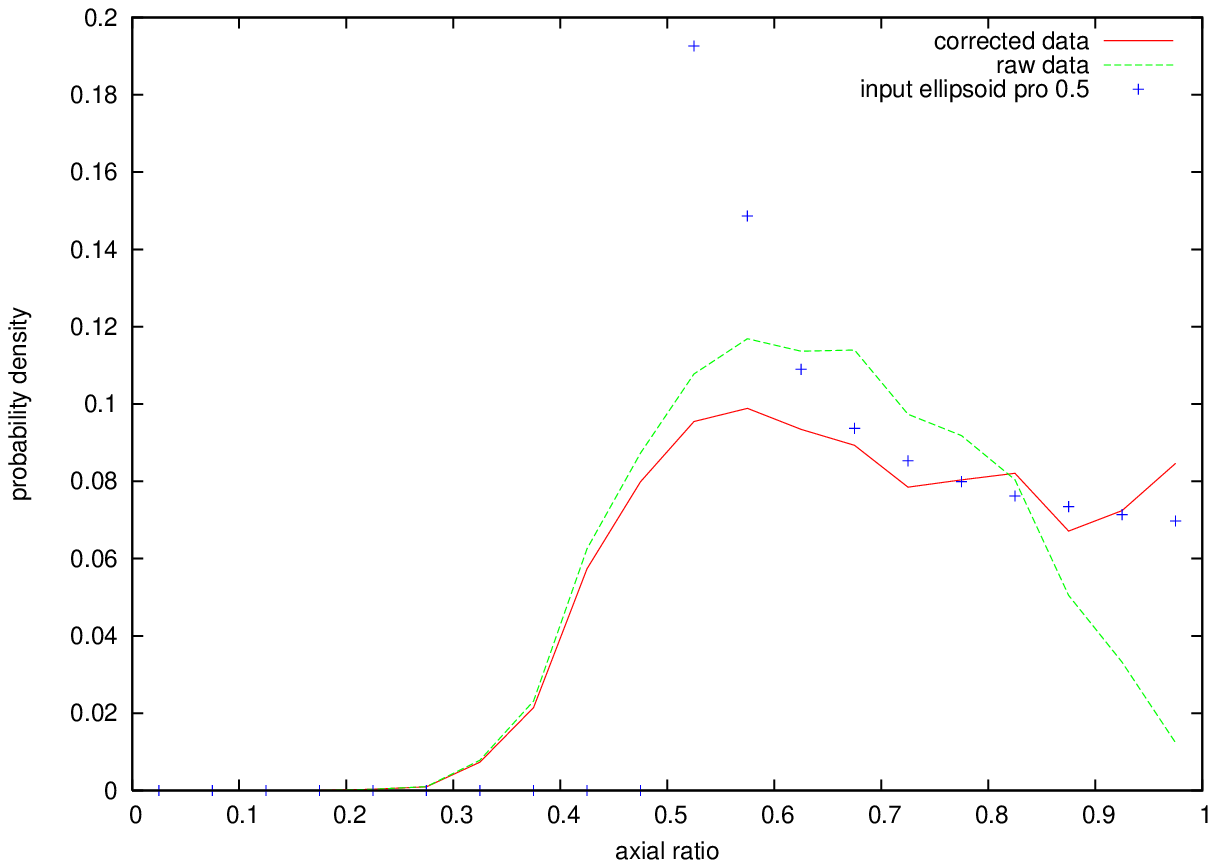}
  \caption{Raw and corrected PDF plotted against the input ellipsoid data for a given group using the HEALPix method to sample the rotation parameter space evenly. On the left is a prolate group of 30 objects with an axial ratio of 0.5. On the right is an oblate group of 10 objects with an axial ratio of 0.5.}
  \label{montellip}
\end{figure*}

So far we have considered the correction required for each individual observation, i.e. the most likely intrinsic ellipticity for a particular observed ellipticity for a certain multiplicity. In practice the rather complex distortions for axial ratios near unity are best accounted for by considering the amount of distortion between the input and measured axial ratio distributions. Using this Monte-Carlo data it is trivial to find the correction factor between a uniform PDF of input axial ratios and the measured non uniform distribution for each multiplicity (see Fig. \ref{pdfax}). This technique has the advantage of correcting for the lack of axial ratios $\sim1$ in low multiplicity groups, and we don't lose any data, it is merely weighted depending on the significance of the measurement. This makes intuitive sense because if we had 10 low multiplicity groups, 5 of which had axial ratios near 1 and 5 have small values, overall the correct interpretation would be that the groups have close to circular true axial ratio since these are the axial ratios that are suppressed (low values of the PDF), and thus carry more significance when actually found. Figure \ref{montellip} demonstrates the effect of this correction on a generated distribution of points for both an oblate and prolate ellipsoid with differing multiplicities. For the prolate group of 20 objects with an axial ratio of 0.5 the best fit was with the corrected data, returning a prolate fit with mean axial ratio of 0.50 (reduced $\chi^2$ of 6.25 compared to 27.5 for the uncorrected data). For the oblate group of 20 objects with an axial ratio of 0.5 the best fit was with the corrected data, returning an oblate fit with mean axial ratio of 0.48 (reduced $\chi^2$ of 2.04 compared to 63.37 for the uncorrected data). In the later case the result was particularly significant because the oblate fit was strongly rejected by the raw data alone, despite us knowing that this should be the returned best fit distribution. These results are encouraging since the convolution correction is best applied to a large amount of input data, such as the catalog data that we have available. The caveat when using this technique is that extremely sharp features will always be smoothed out in low multiplicity data since the axial ratios themselves are not re-binned, only re-weighted; this effect being obvious in \ref{montellip}. The type of distribution and the mean are reliably recovered however, and this is the the matter of most importance. 

These two different techniques, simply correcting the means of the distributions for sampling effects or using the whole PDF, can be used to produce two different distributions of axial ratios. When used on test data they display good agreement for large multiplicity samples; divergence at large axial ratios is always present however, since for any multiplicity the Monte-Carlo tests strongly suggest axial ratios near 1 are observationally inhibited. This is significant since it is the axial ratios near 1 that help us to distinguish between prolate and oblate underlying distributions: oblate distributions plateau at large axial ratios, whilst prolate distributions will drop to extremely low values.

Considering the first of these two methods for creating our PDFs, a cut-off was applied to the sample, requiring a minimum of 20 galaxies in order to achieve an acceptable level of confidence in the $\varepsilon$ values returned. In addition, a lack of sufficient multiplicity will always have the effect of increasing the ellipticity measured; the obvious extreme example would be a multiplicity of 2, where the measured ellipticity will always be 1, no matter the distribution that created them. Obviously, the ideal situation would be to increase the multiplicity limit further, and it is true that this would increase the confidence in our measured ellipticities, but two factors militate against this: first we are interested in a range of group multiplicities, so would like to consider as broad a sample as possible, and second we need as many measured ellipticities as possible in order to produce a well defined binned PDF $(N(q))$. The remaining groups were analysed in the same fashion as the Monte Carlo models and sorted into axial ratio bins of size 0.05 from 0 to 1, creating a binned PDF that can be compared to the distributions predicted for the models in Section 2.

For the second method all group multiplicities are used, the only limitation being those of the original data set; i.e. multiplicities of 5 or more. A convolving routine was written in order to weight the significance of each result, which is a combination of the multiplicity and the axial ratio measured. These results were then placed in the same bins as used for the previous technique, allowing for direct comparisons. To aid analysis Monte-Carlo distributions were also created, these correspond to the PDF that would have been observed had the true PDF of axial ratios been uniform for each group multiplicity.

\begin{figure*}[t]
\plottwo{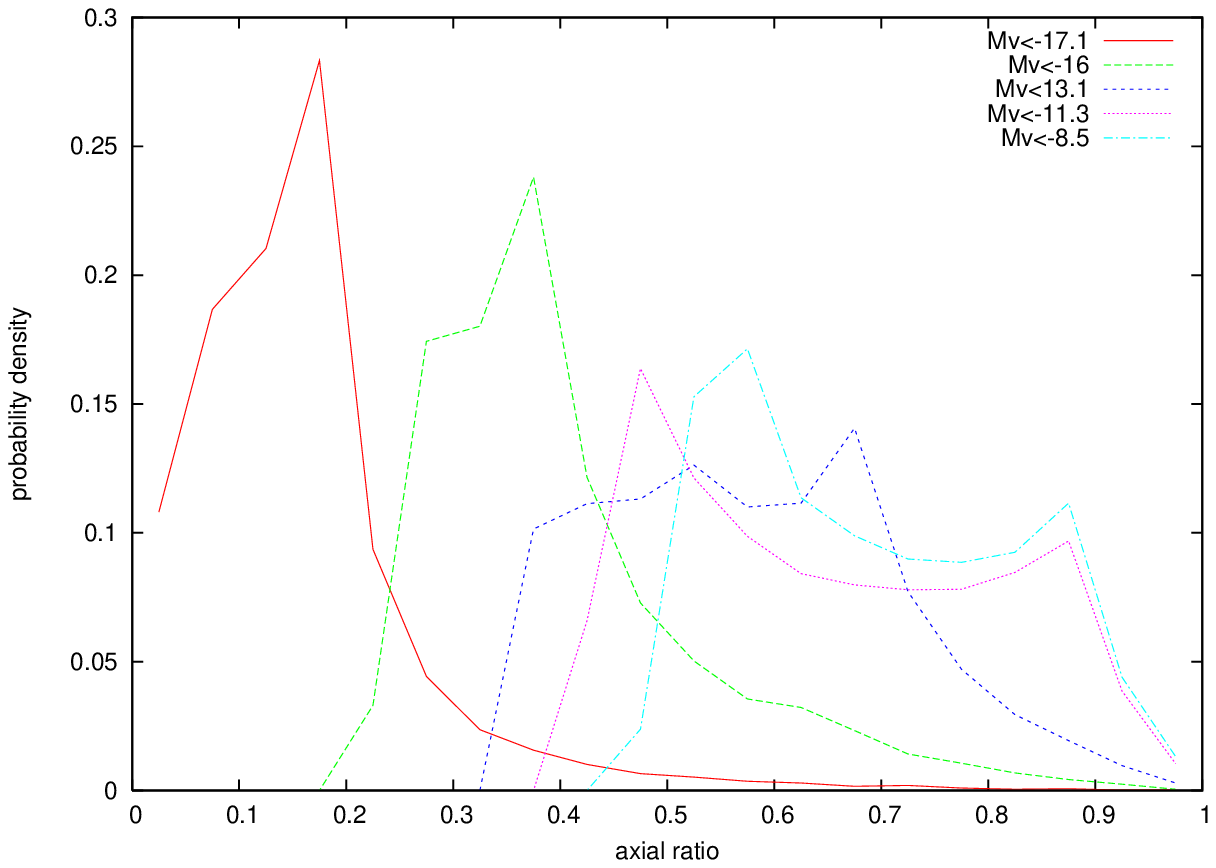}{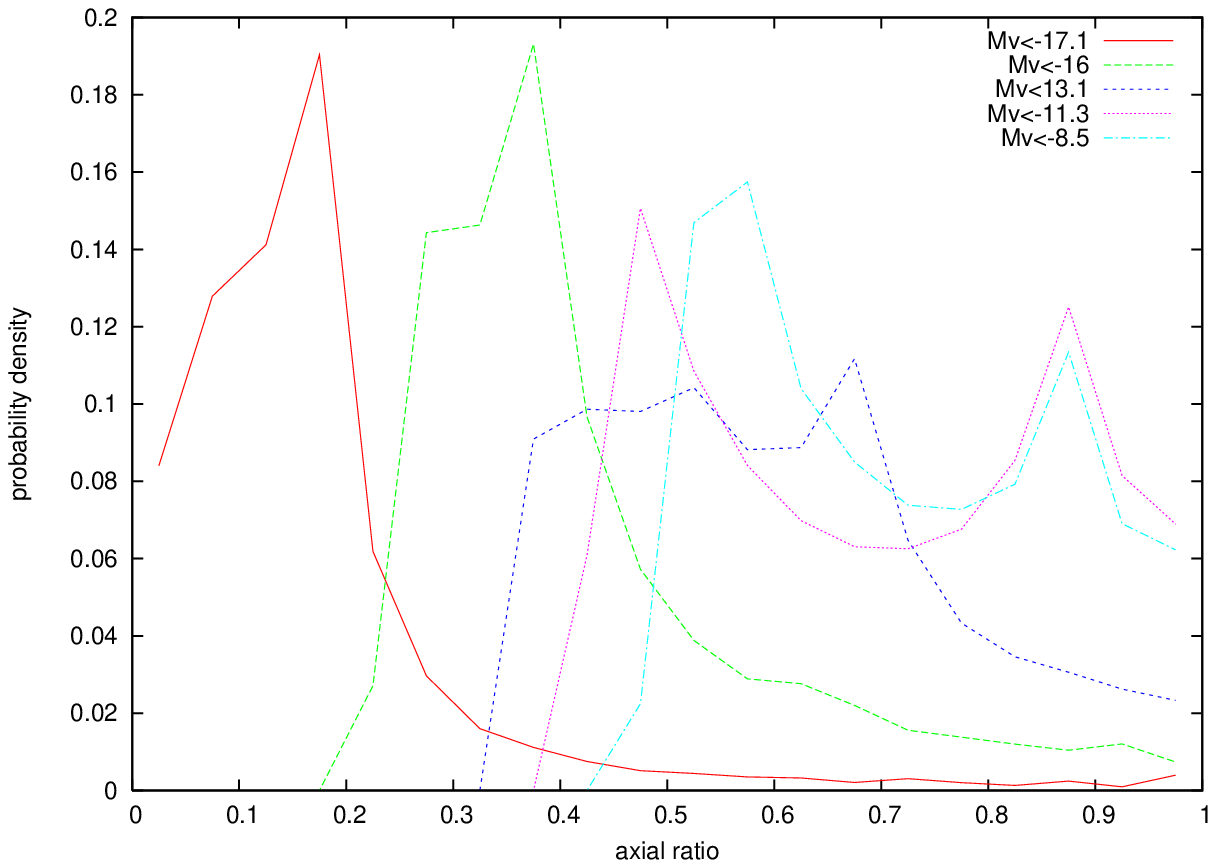}
  \caption{Local group projected axial ratios at different V magnitude cuts using the HEALPix method to sample the rotation parameter space evenly. On the left are the uncorrected distributions. On the right are the convolution corrected distributions.}
  \label{localgroup}
\end{figure*}

As a point of comparison figure \ref{localgroup} shows how our Local Group appears projected on the sky at different magnitude cuts (data taken from \citealt{pritch99}), evenly sampled using the HEALPix isolatitude pixelization method to ensure a representative distribution of convolution corrected and non corrected projections of the Local Group in our rotation parameter space. The effect of the correction is most evident on the low magnitude cuts where the plateau at larger axial ratios is suppressed in the raw data, but after correction the plateau continues more evenly. This suggests the Local Group is oblate at low magnitude cuts. The extremely small axial ratios at high magnitudes is understandable in the context of the Local Group having significant substructure, and at high magnitude cuts only the regions around the Milky Way and M31 contribute. This creates a dumbbell structure that is increasingly smoothed out as fainter members are added, hence the systematically larger axial ratios as the magnitude limit is increased.

\section{Contamination by Interlopers}

A problem for any attempt to collect galaxies into associated groups and clusters is the unavoidable presence of interlopers, as such it seems prudent to quantitatively define what the effect is. The situation is improved by the use of spectroscopy since this requires interlopers to have chance projections and chance recessional velocities, however it is clear this still won't eliminate all interlopers, just reduce the percentage of the population that are false. The 2PIGG catalog is spectroscopically selected, so it is as good as reasonably possible. Figure 4 of \citet{eke04} shows three plots to describe the typical interloper rates in groups of different mass, redshift and multiplicity. As might be expected, interloper rates get worse with the largest multiplicities (hence mass) and at the highest red shifts, however the redshift cut we have applied at 0.1 indicates a very steady interloper rate of 20\% to 30\%. Over our regime of group masses ($\sim10^{13}M\sun$ to $\sim10^{14.5}M\sun$) and multiplicity (5 to 163) the 2PIGG catalog is also predicted to maintain a consistant interloper rate.

Since interlopers are an inevitable feature of our catalog it is of paramount importance to discern what effect they have. Assuming a given group has an intrinsic shape seen in projection, the consequence of interlopers will be to add a circularly distributed population on top of any projection of the same maximum allowed radius. This is obvious since any grouping algorithm merely considers the magnitude, radial separation and, possibly, redshift of any potential new member. Intrinsically no restriction is made as to how objects affect the shape of the currently grouped objects. With this in mind a Monte-Carlo simulation was designed to find how mixing together inrinsically shaped ellipsoids of oblate and prolate varieties with an interloper population can distort the observed results. Of particular interest is how the same rate might affect different multiplicities.

For the sake of interest a large variety of interloper rates were investigated, from $2\%$ to $50\%$, although our main interest will be the interloper rate $\sim20\%$. Multiplicities of 5, 10, 20, 50 and 200 were chosen to allow for population comparisons, and axial ratios of 0.05 upto 0.95 were generated. For each Monte-Carlo distribution (a given interloper rate, multiplicity, ellipsoid type and axial ratio) 20,000 virtual groups with interlopers and 20,000 groups without interlopers were generated (the number of runs was deemed more than adequate in pre-tests). KS-tests were carried out between all populations of interloper and non-interloper groups and the best fit solution determined. The result of this is a direct measure of how interlopers can distort what is observed, and can be used as a transformation guide to find the real shape of the underlying ellipsoid.

\begin{figure*}[t]
\plottwo{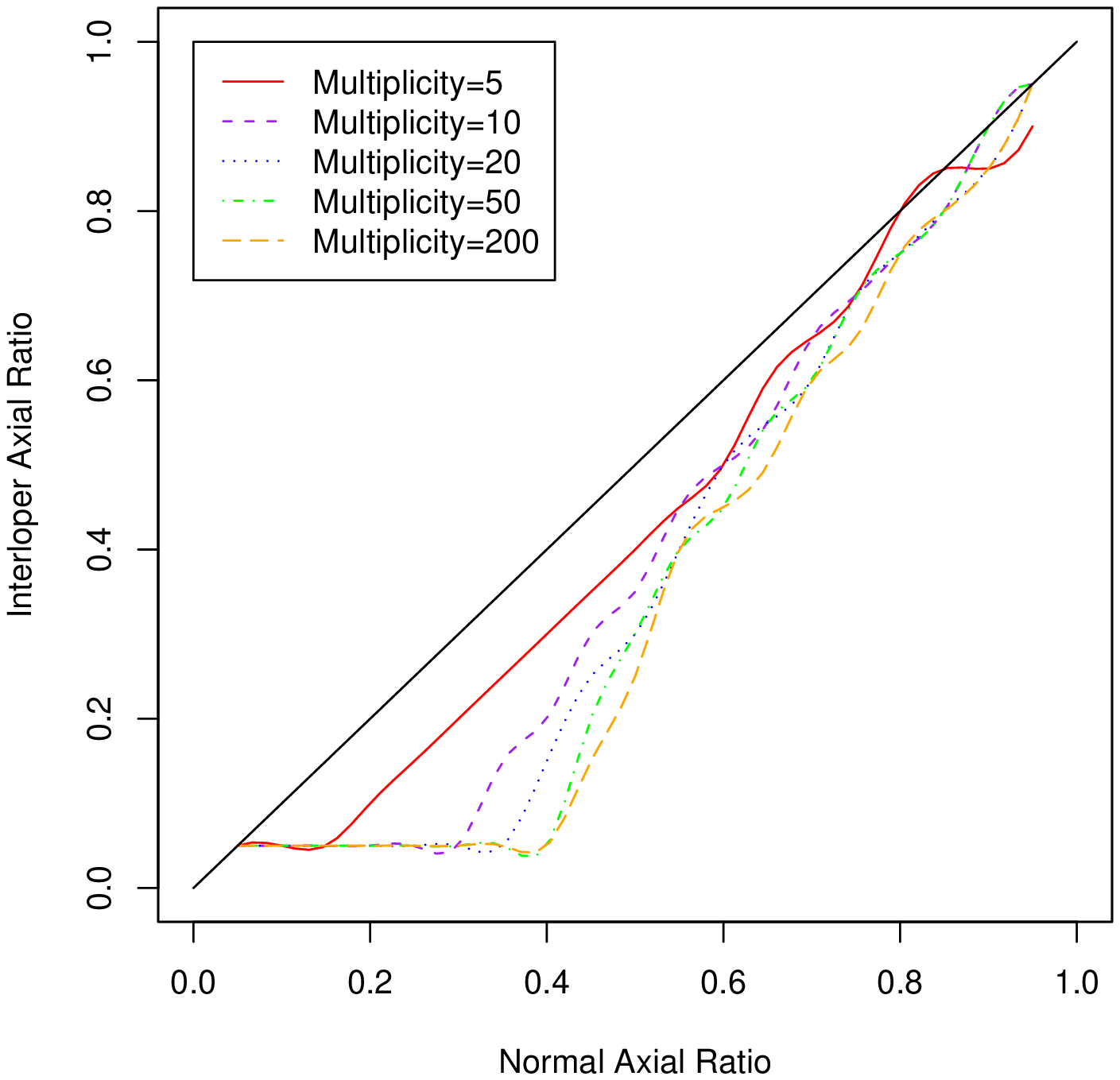}{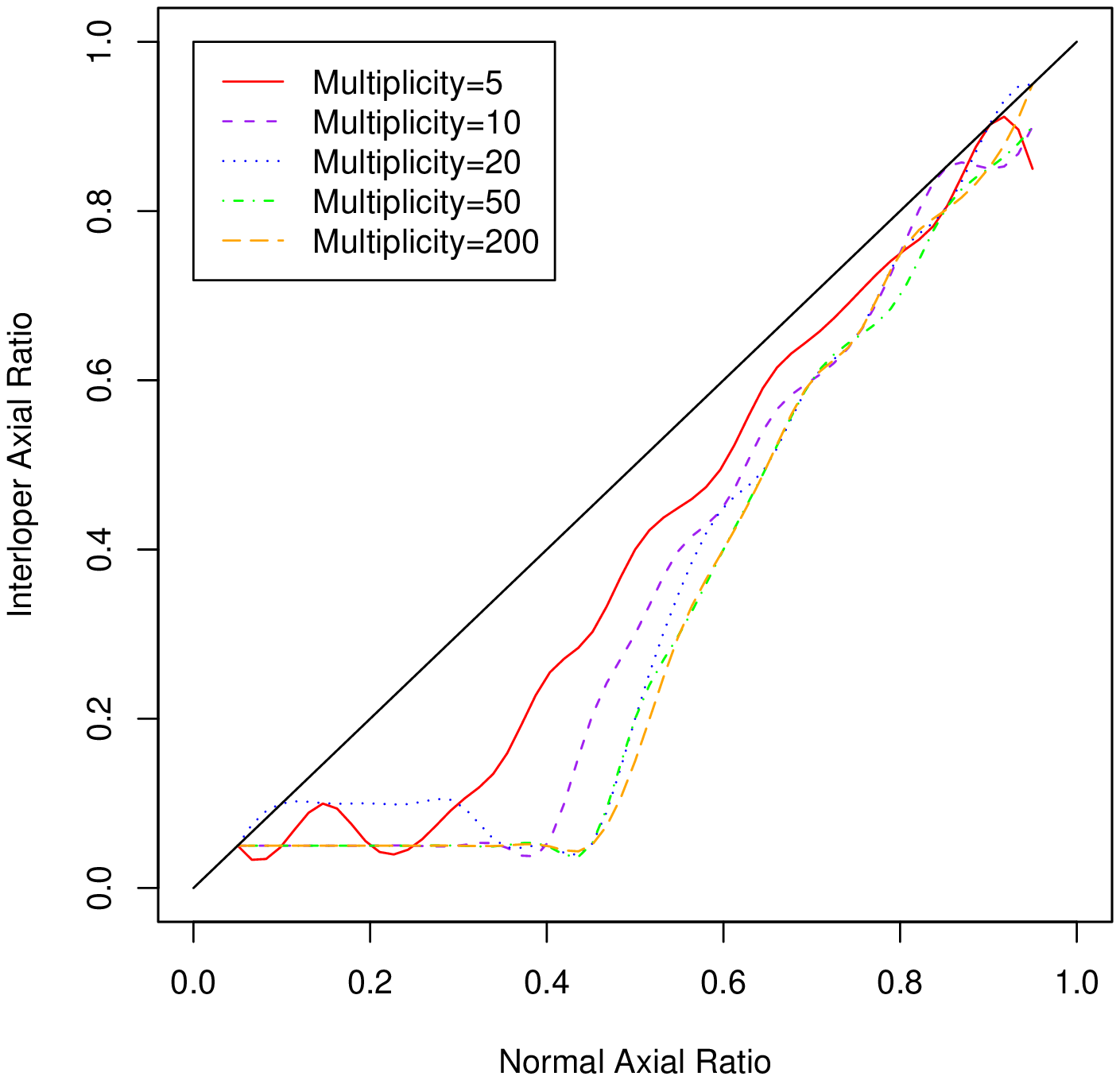}
   \caption{The effect a 20\% interloper rate has on prolate (left) and oblate right) best fit distributions.}
  \label{interplots}
\end{figure*}

\begin{figure}[t]
\plotone{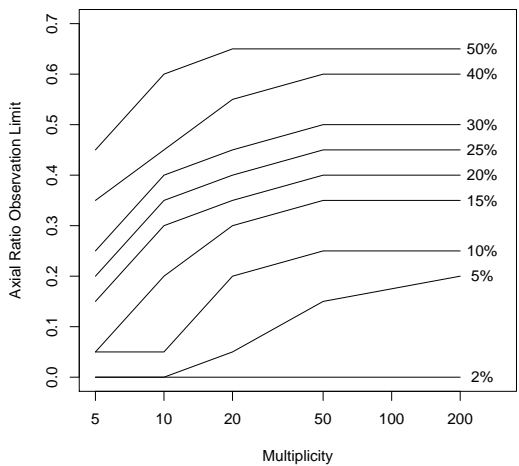}
\caption{Absolute detectable limits for given multiplicities and interloper rates.}
\label{limits}
\end{figure}

\begin{figure*}[t]
\plottwo{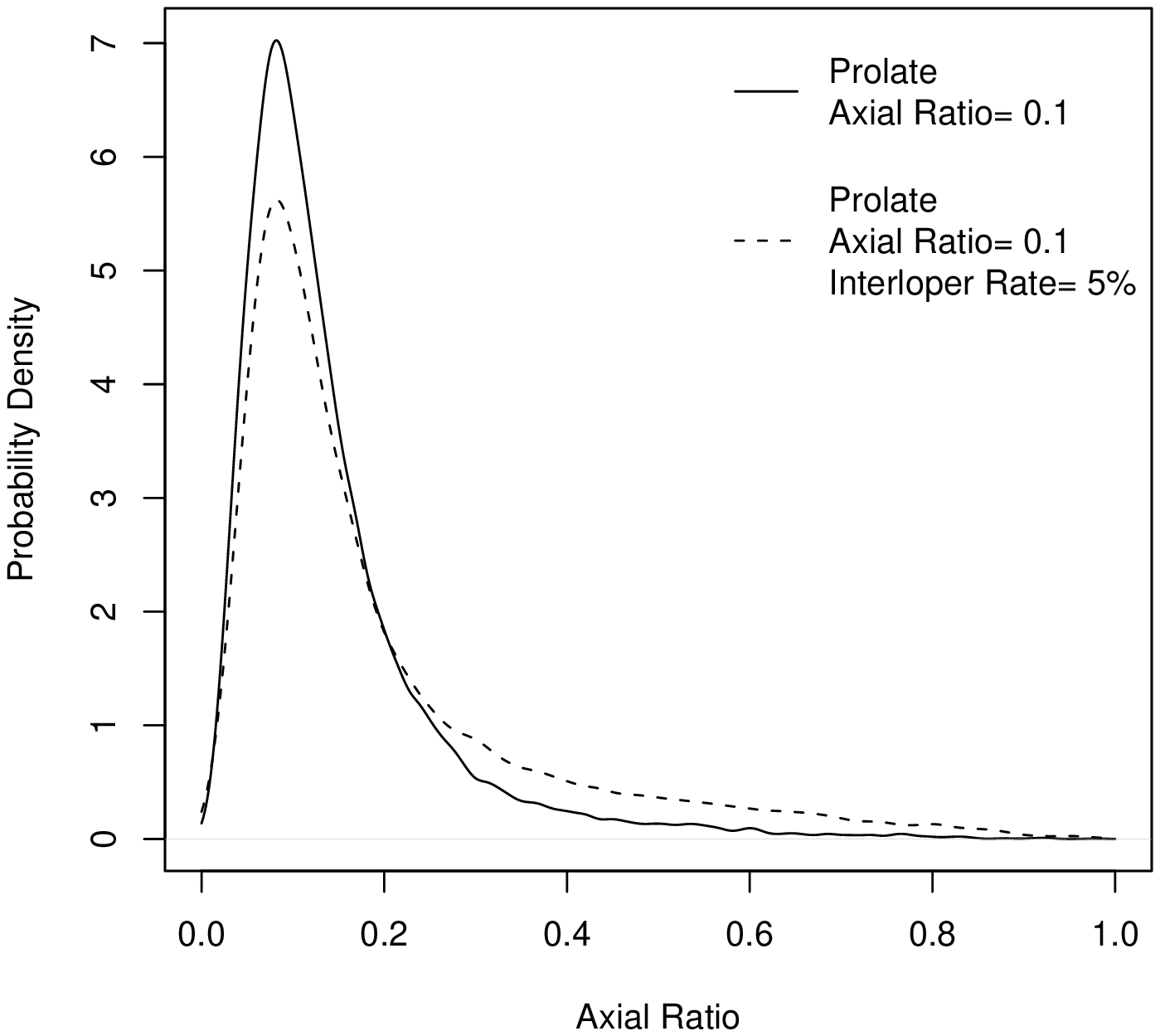}{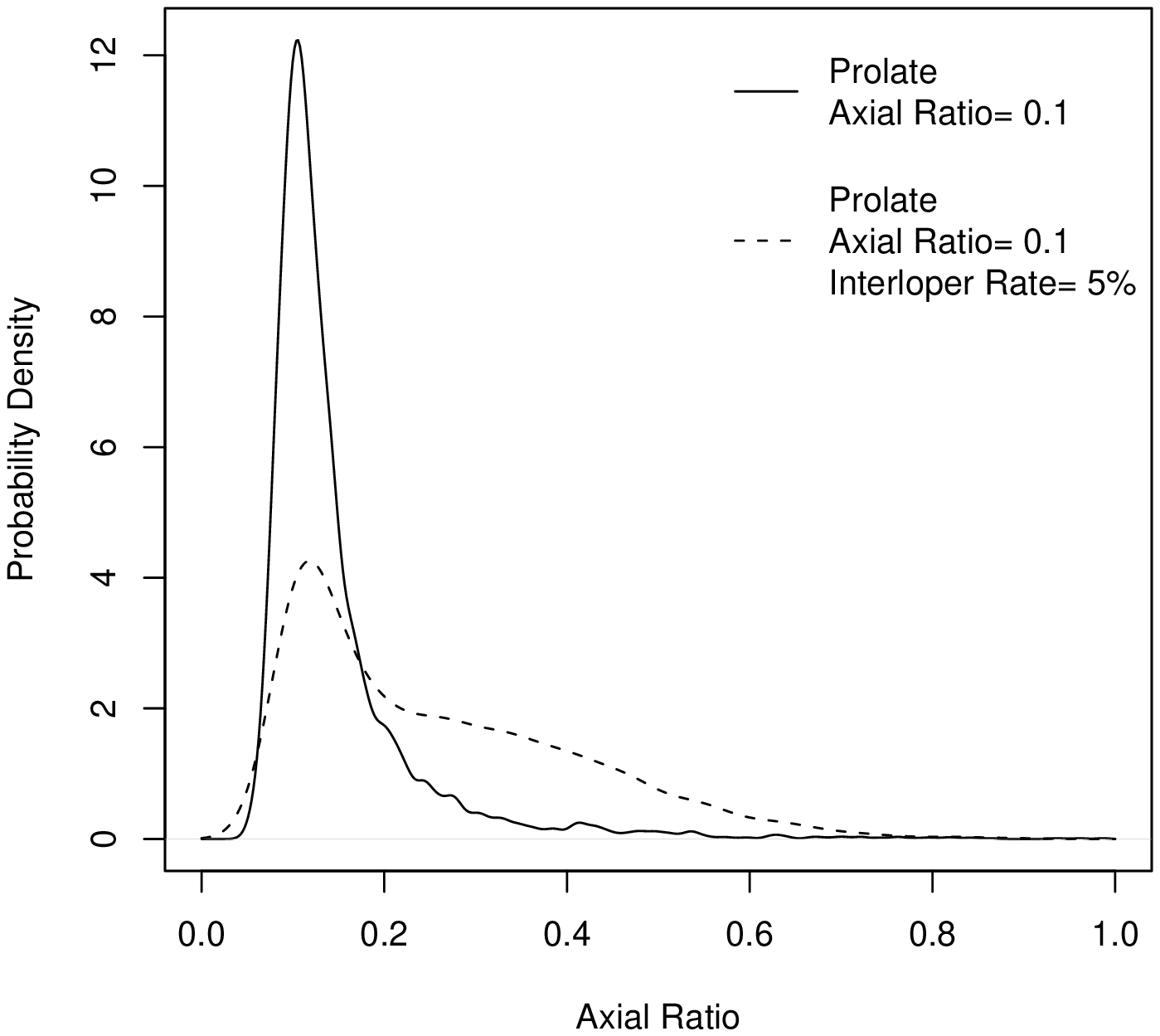}\\
\plottwo{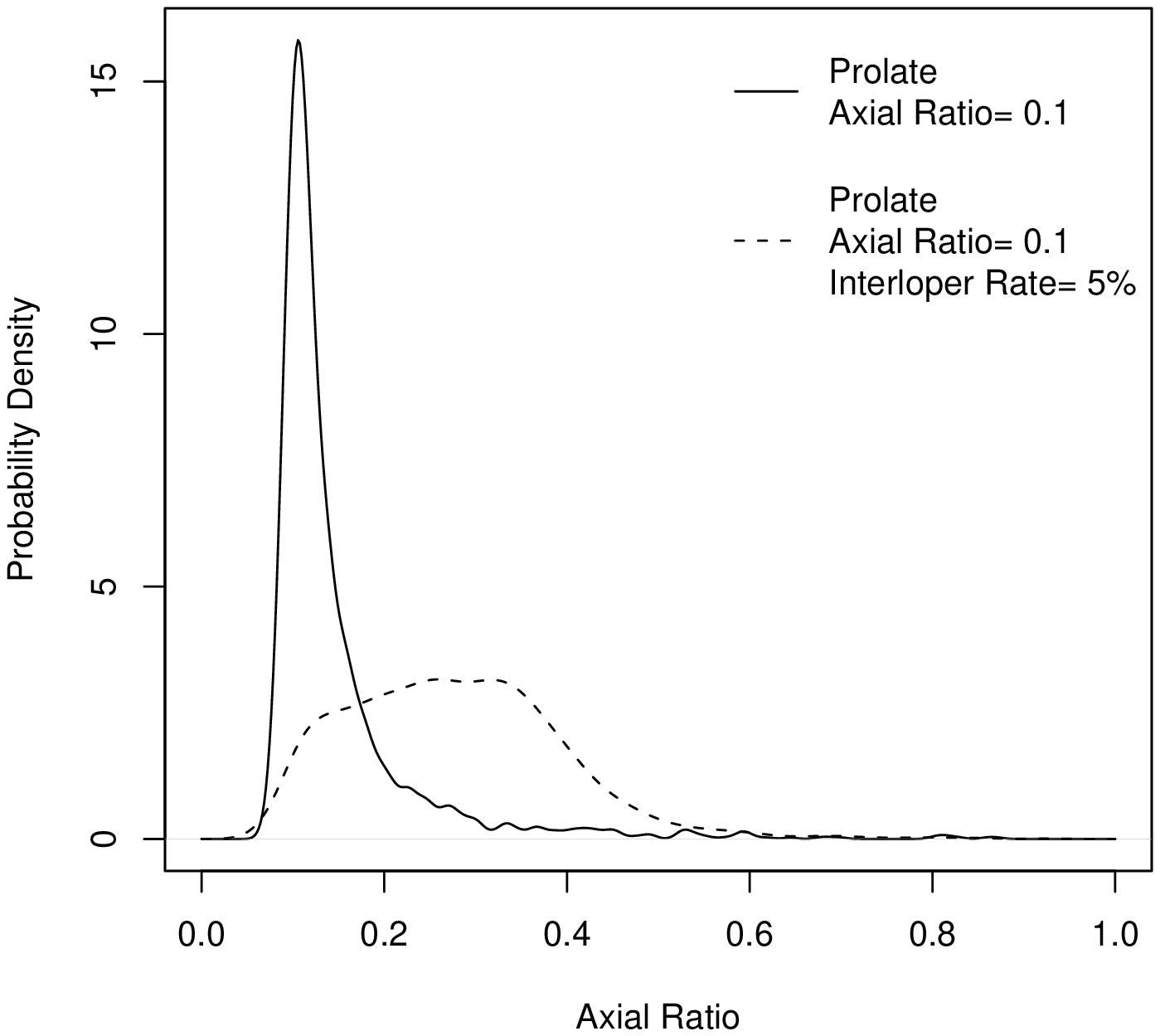}{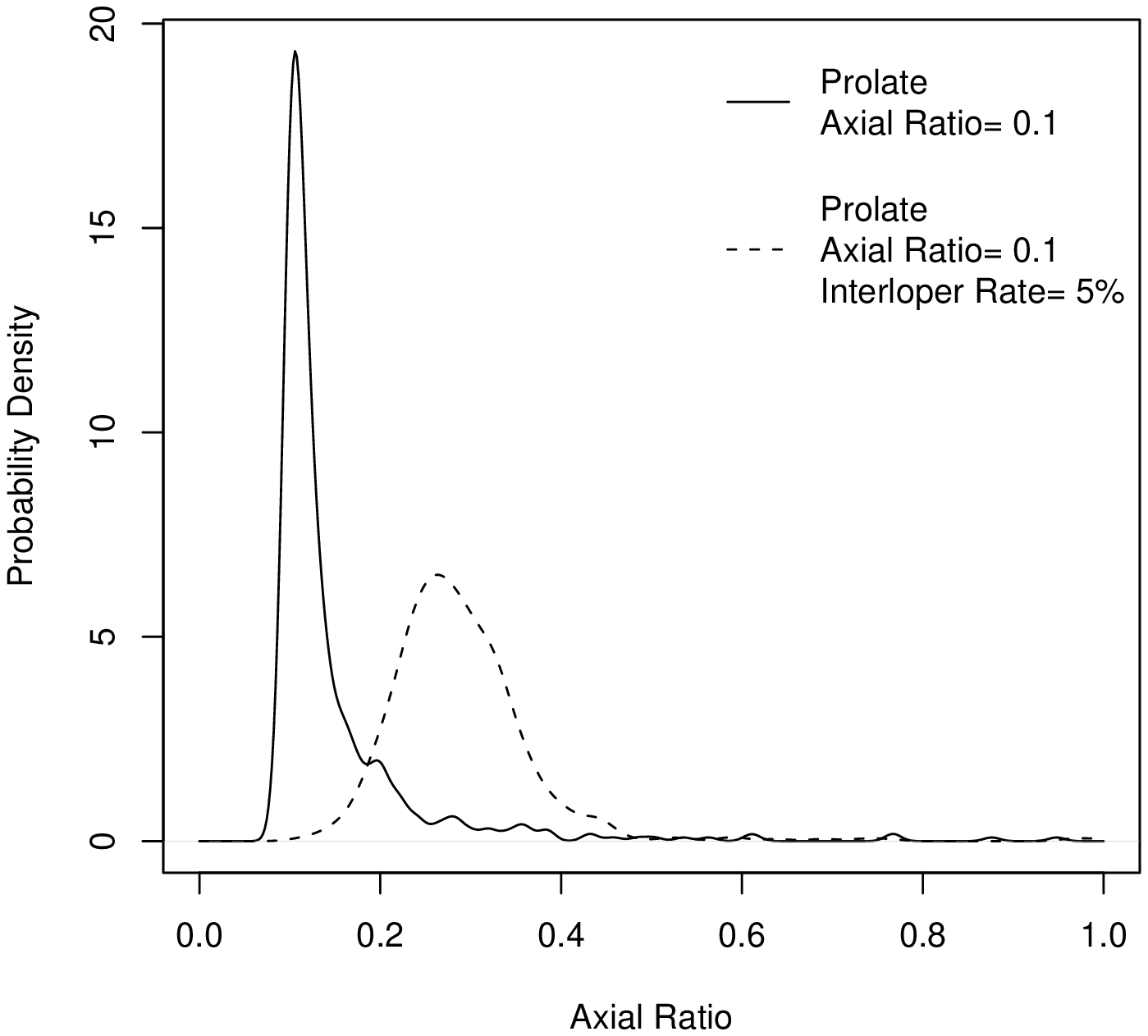}
  \caption{Distortion casued by 5\% interloper rate for prolate groups of multiplicity 5 (top-left), 20 (top-right), 50 (bottom-left) and 200 (bottom-right).}
  \label{distortions}
\end{figure*}

\begin{figure*}
\plotone{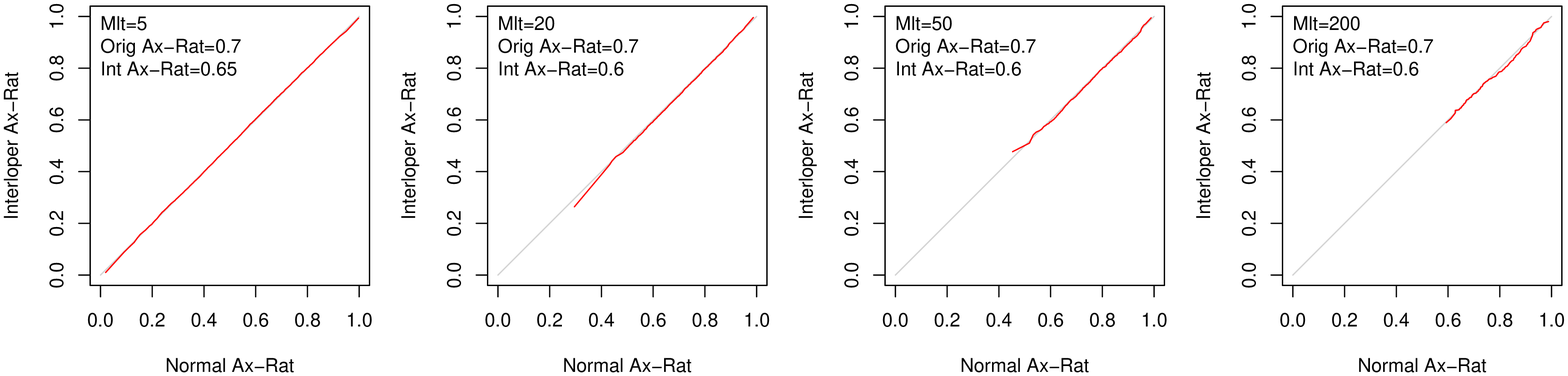}
\caption{QQ-plots for best fit interloper groups where interloper rates= 20\% and axial ratio= 0.7.}
\label{qqplots}
\end{figure*}

The main finding is that for all axial ratios the interloper groups are consistantly observed to be more spherical than their underlying population. As might be expected, the effect is weakest for the lowest interloper rates and vica-versa (see Figure \ref{interplots}). Perhaps the most interesting aspect to note is that the lower the multiplicity the smaller the distortion to the interloper distribution compared to the normal one, whether oblate or prolate. Another interesting outcome of these simulations is we find a hard limit for the observable ellipsoid as a function of multiplicity and interloper rate (see Figure \ref{limits} for the prolate limits). The observable limit is smallest for groups with the lowest interloper rates and multiplicities, and worst for those with the highest interloper rates and highest multiplicities. The dependence on interloper rates is how one might expect, but the relation to multiplicity might not be so obvious. These trends are observed strongly for both prolate and oblate shapes.

A significant issue is the confusion effect: does the presence of interlopers make a prolate distribution look oblate and vica-versa, and how does multiplicity and the interloper rate affect the possibility of confusion? When confusion occurs it is generally at the largest axial ratios, an understandable result since as prolate and oblate shapes move closer to spherical it requires a smaller amount of random distortion to transform one distribution into the other, and hence it doesn't affect prolate or oblate groups worse.

It becomes easier to distinguish populations when the axial ratios are lower ($\sim0.5$), however this improvement is lost when axial ratios are very small (less than the effective observable limit) because the distributions become very hard to fit at all. The distortion is particularly evident for higher multiplicities, as can be seen in Figure \ref{distortions}. Here all the input axial ratios are 0.1 and the interloper rate is only 5\%, and yet the amount of distortion becomes quite significant as a function of multiplicity. In contrast the quality of fitting for axial ratios of 0.7 is very good even when interloper rates are at 20\%, as is evident from the QQ-plots in Figure \ref{qqplots}.

The last finding of note is that interloper rates alone account for some degree of distribution broadening, on top of the aforementioned shift to more spherical populations. This is understandable simply because we are modeling a binomial distribution (every Monte-Carlo galaxy has a chance of being an interloper as described by the interloper rate), and as such there will be an associated spread in observed shapes since when there are many interlopers a given group will appear more circular and when there are fewer this effect is reduced. In fact the number of interlopers will follow a near gaussian distribution when {\it np}$\sim 5$, so groups with multiplicity 20 and interloper rates of 20\% almost meet these conditions. Otherwise we'll still observe a spread, just of a more discrete variety. This effect is clearly seen in Figure \ref{distortions} where only the multiplicity 200 population undergoes a smooth distortion; this is because {\it np}$= 10$ and meets the requirements to approximate a gaussian form for the number of interlopers. The multiplicity 50 population has {\it np}$= 2.5$, so whilst the distribution is broadened it is not as smooth. The multiplicity 20 population has a very strong feature from 0.2-0.5. This is because {\it np}$= 1$ and as such the event of no interlopers is quite likely (36\%), in fact it is almost exactly the same chance as one interloper (38\%). Thus the two main features--the peak between 0-0.2 and the plateau from 0.2-0.5-- are caused by interloper numbers of 0 and 1 respectively. The slight hints of further plateaus beyond 0.5 are caused by interloper number beyond 1, and these only occur $\sim 26\%$ of the time.

These results demonstrate that any observed gaussian distribution of ellipsoid axial ratios are, in general, describing the upper limit of the axial axial ratio and the upper limit for the standard deviation. As the axial ratio is beyond the observing limit it is possible to make an estimate of the underlying distribution, bearing in mind the interloper rate and multiplicity.

\section{Results}

\begin{figure*}[t]
\plottwo{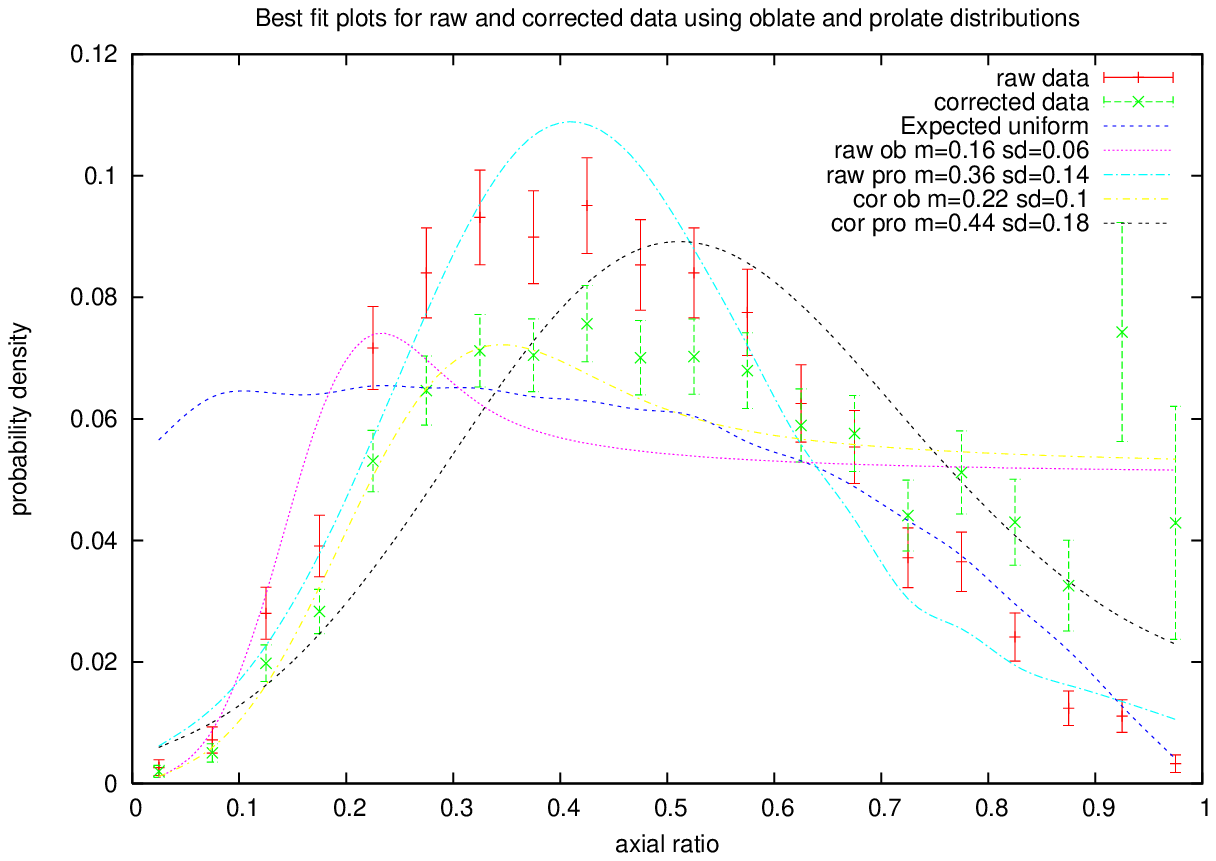}{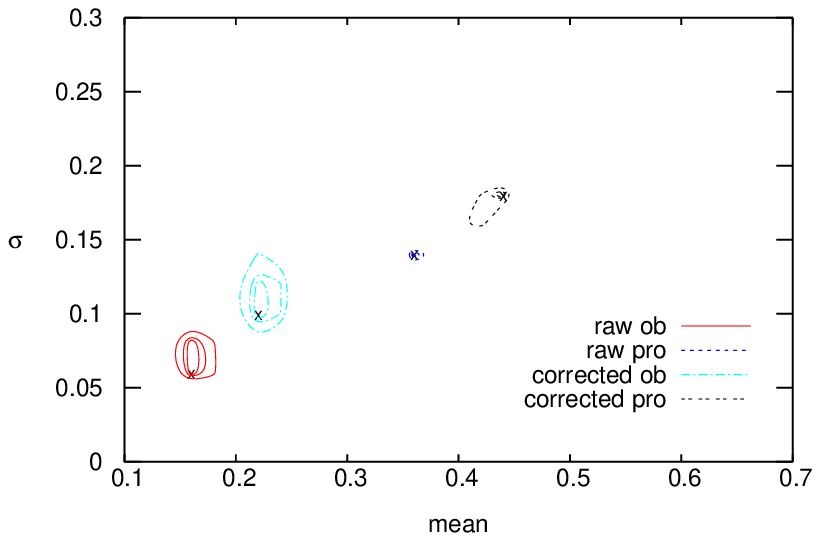}\\
\plottwo{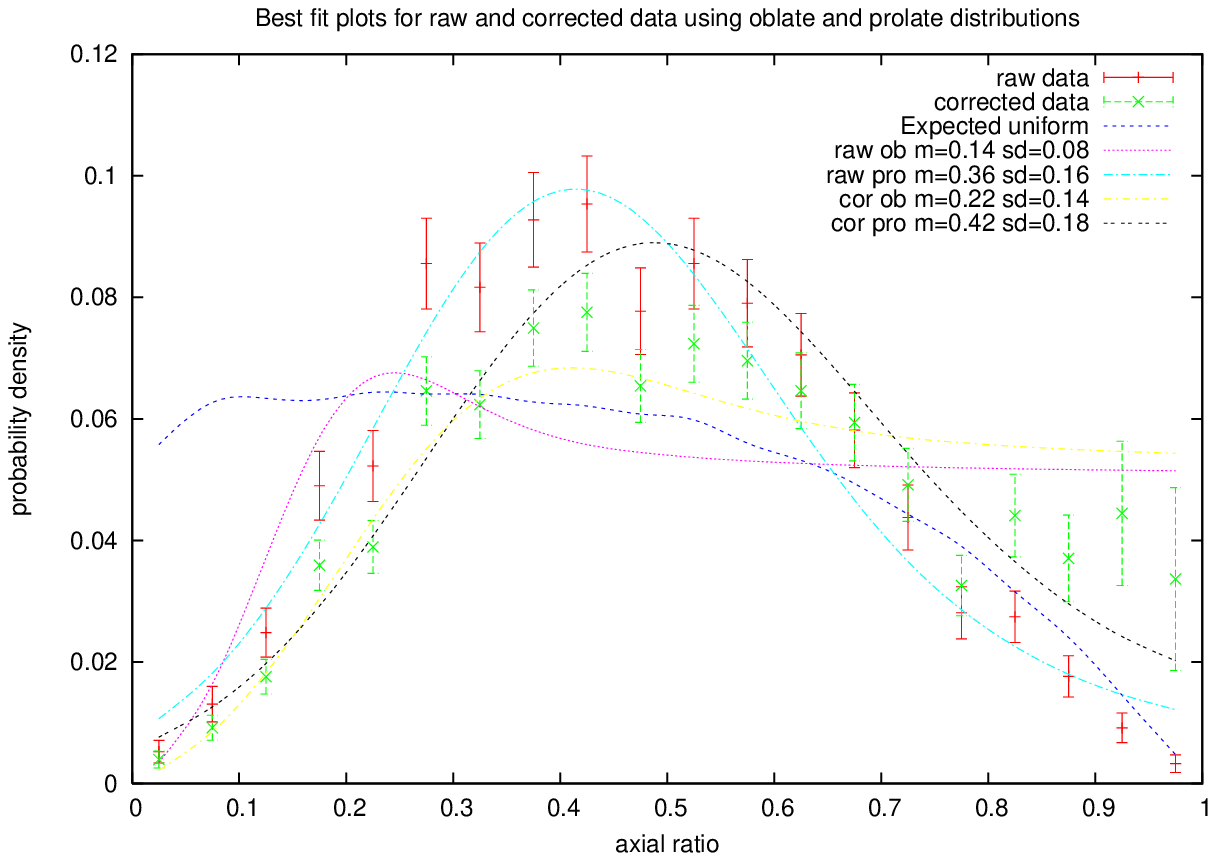}{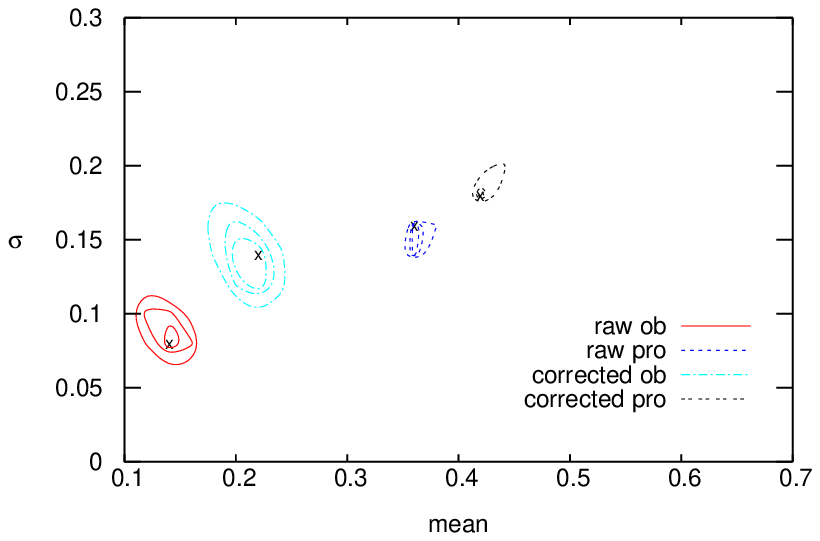}
  \caption{All group multiplicities. Top plot shows best fits for the real 2PIGG data on the left and the relevant parameter contour plots on the right, whilst the bottom plot is a comparison to the mock 2PIGG data.}
  \label{fitschi5+}
\end{figure*}

\begin{figure*}[t]
\plottwo{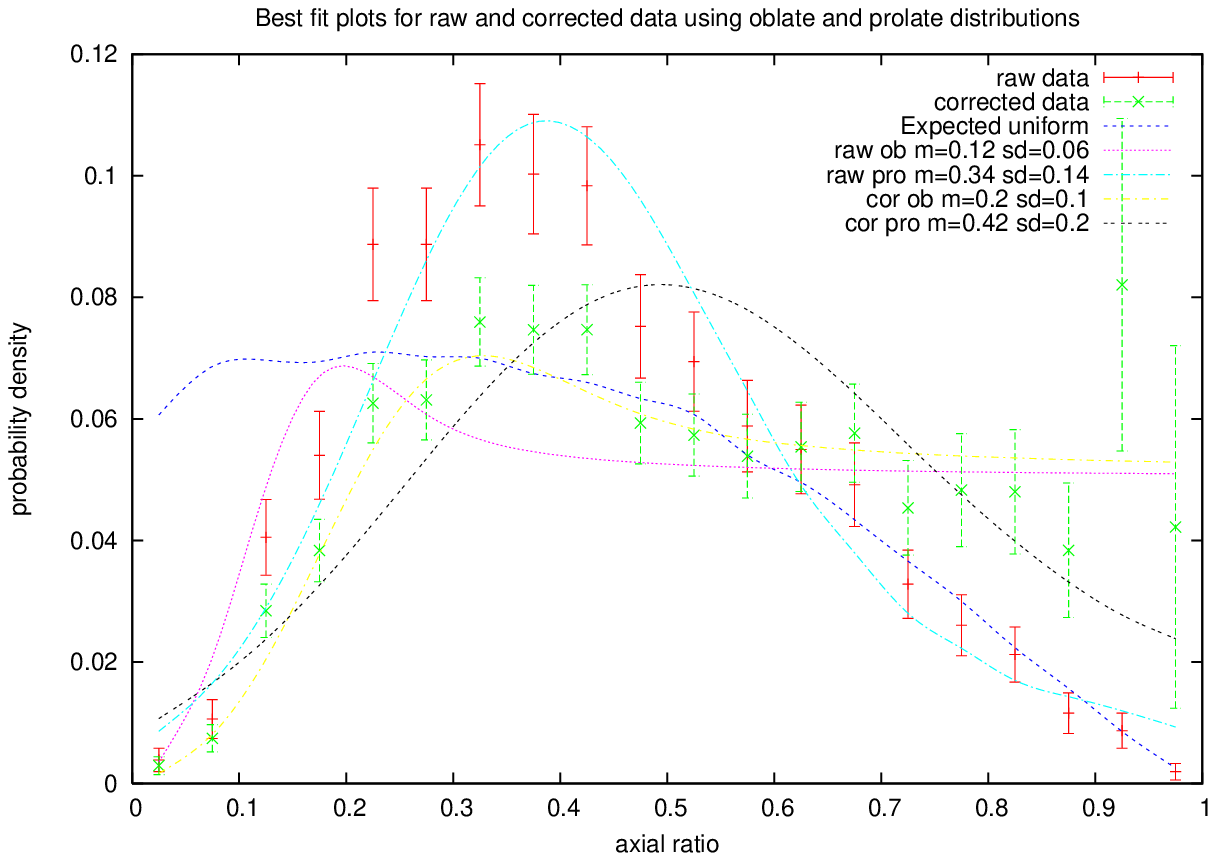}{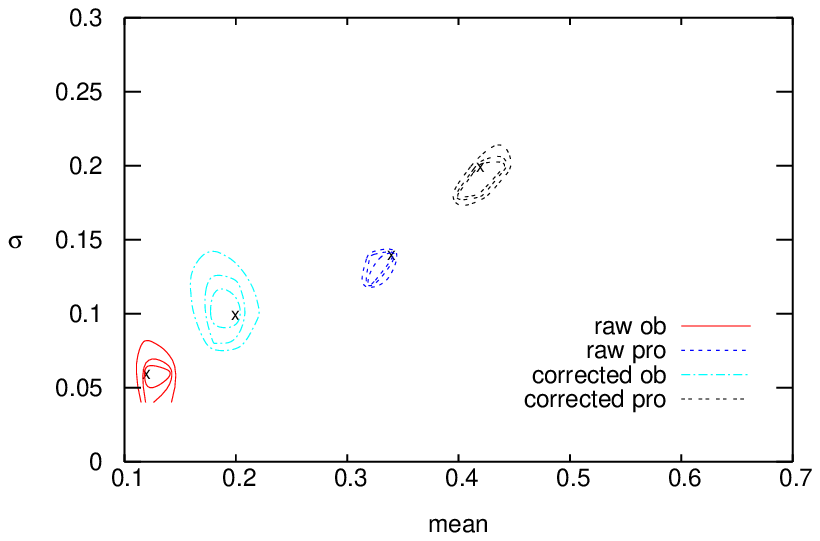}\\
\plottwo{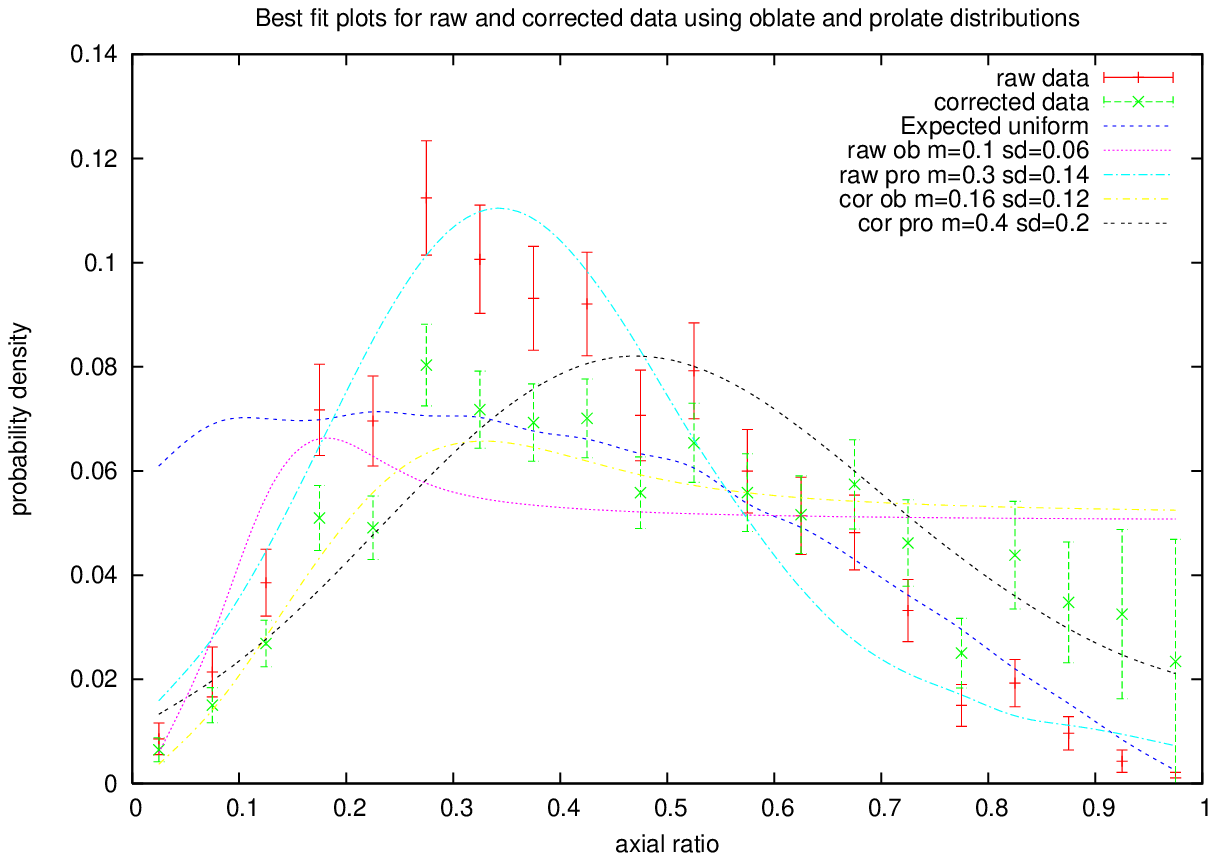}{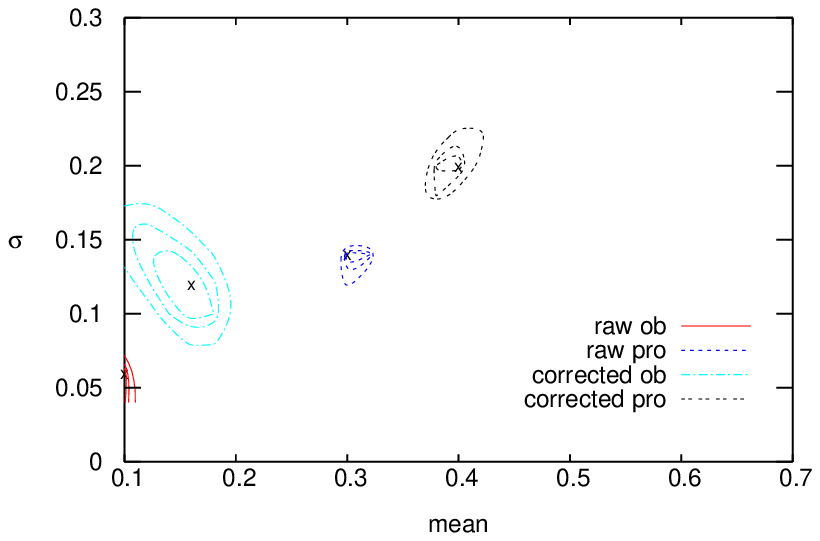}
  \caption{Group multiplicities of 5 to 9. Top plot shows best fits for the real 2PIGG data on the left and the relevant parameter contour plots on the right, whilst the bottom plot is a comparison to the mock 2PIGG data.}
  \label{fitschi510}
\end{figure*}

\begin{figure*}[t]
\plottwo{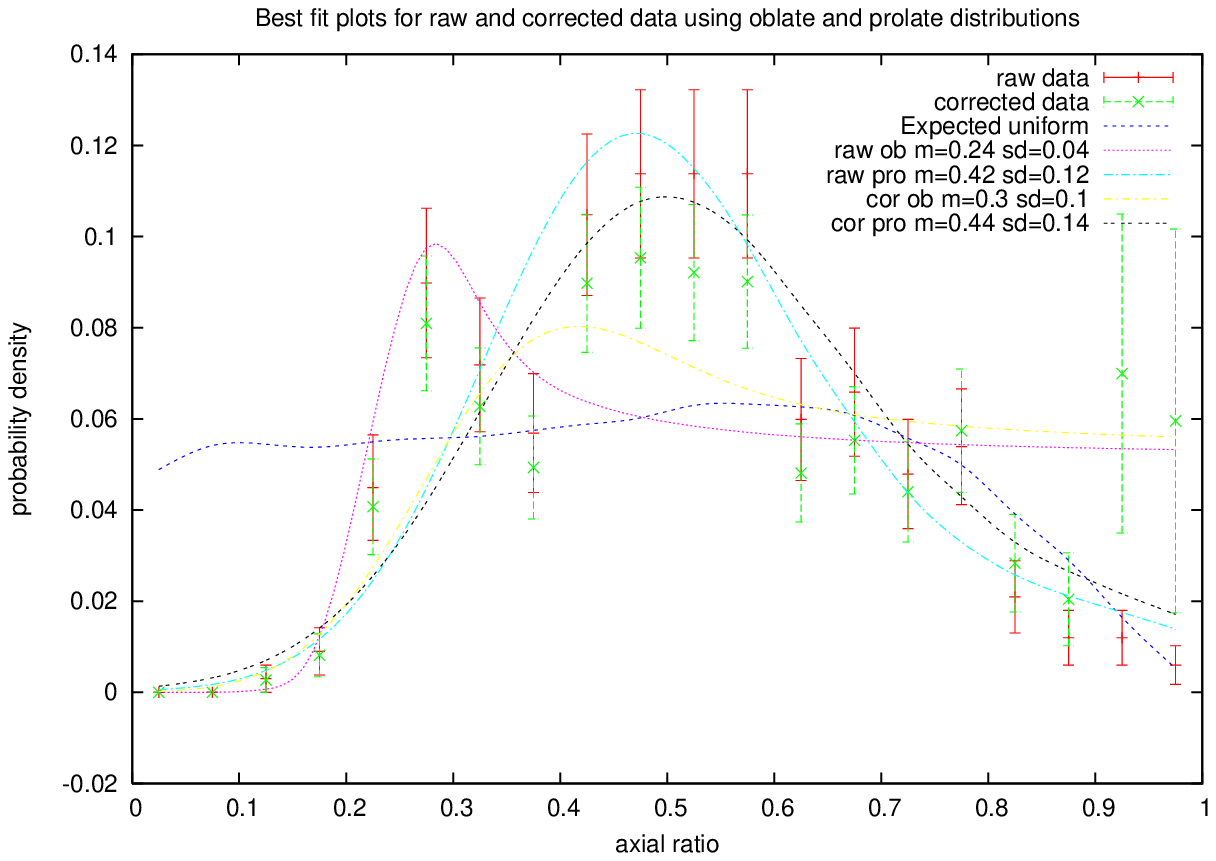}{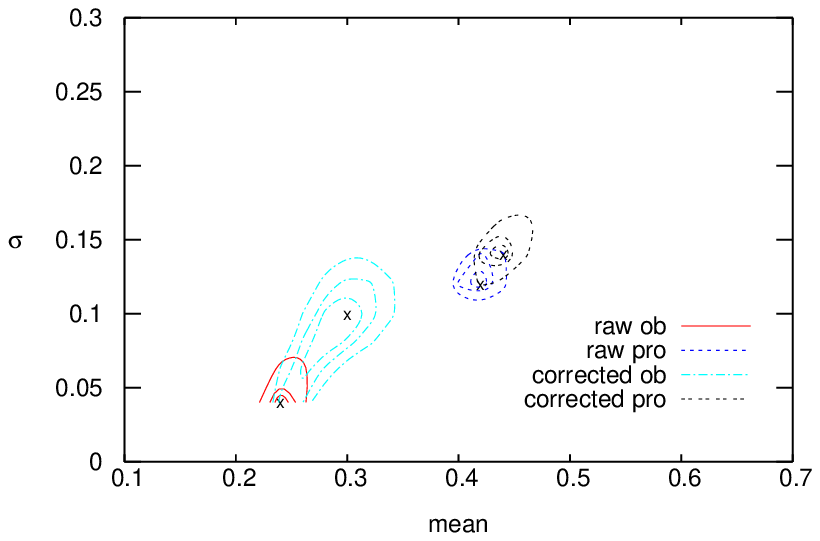}\\
\plottwo{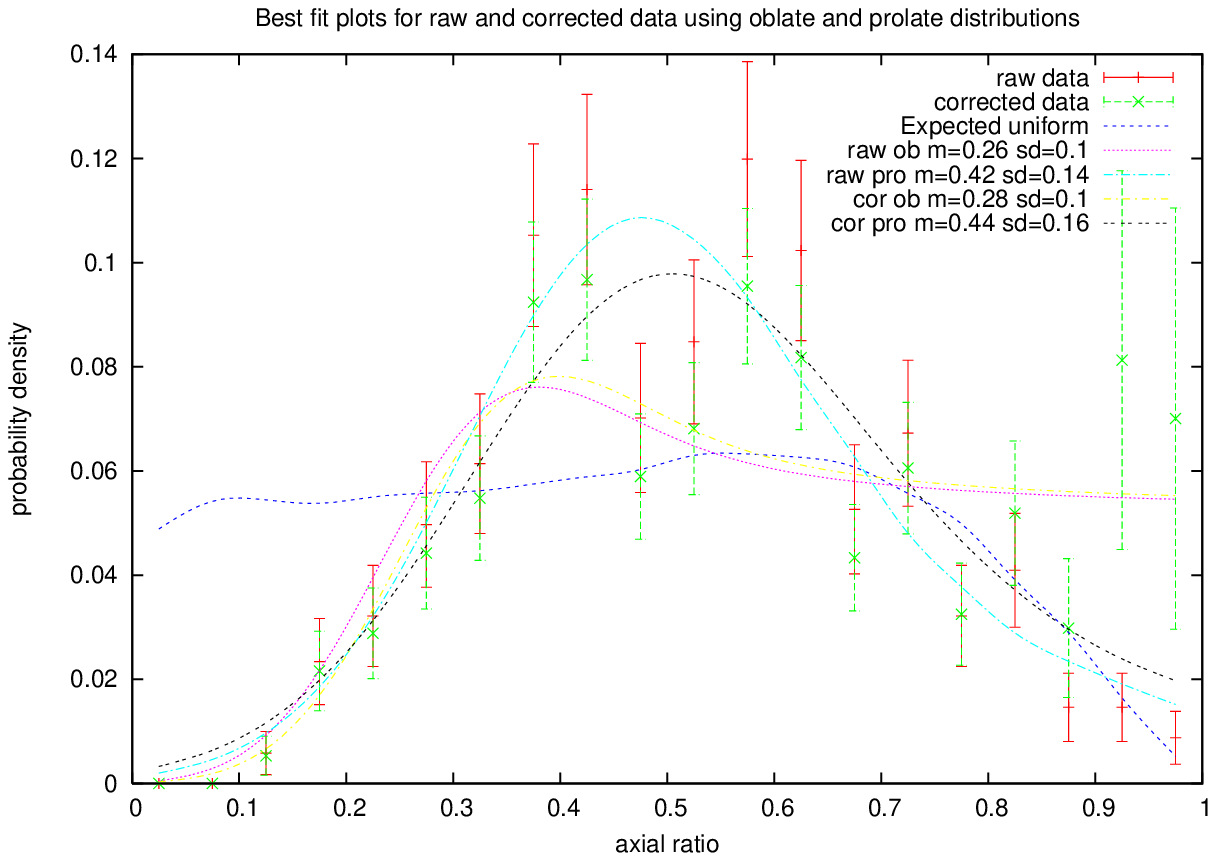}{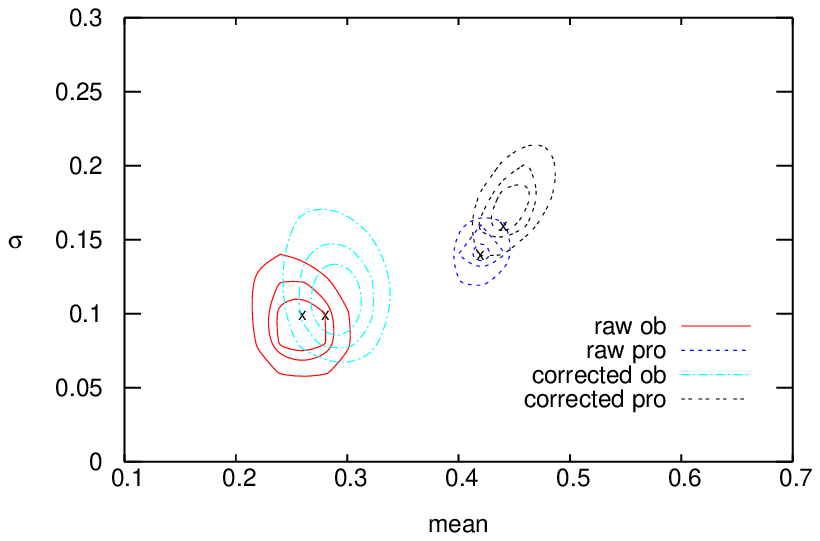}
  \caption{Group multiplicities of 10 to 19. Top plot shows best fits for the real 2PIGG data on the left and the relevant parameter contour plots on the right, whilst the bottom plot is a comparison to the mock 2PIGG data.}
  \label{fitschi1020}
\end{figure*}

\begin{figure*}[t]
\plottwo{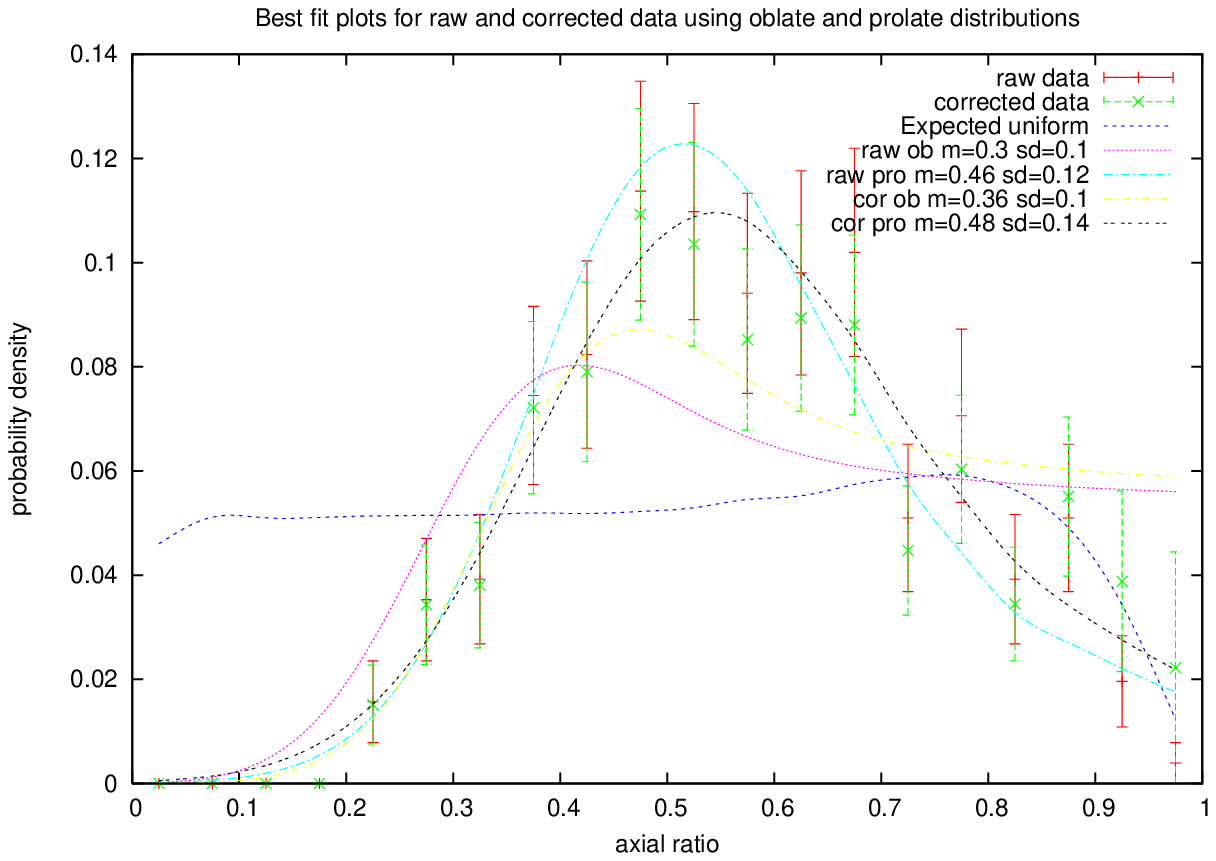}{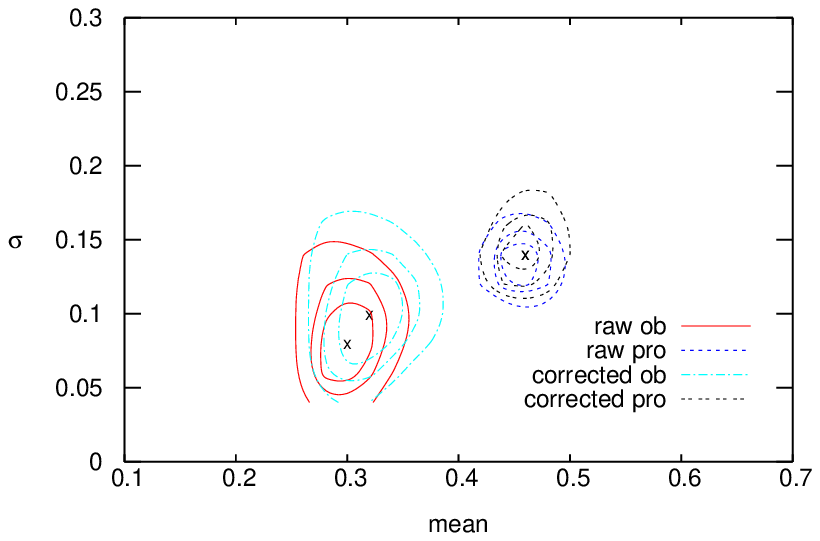}\\
\plottwo{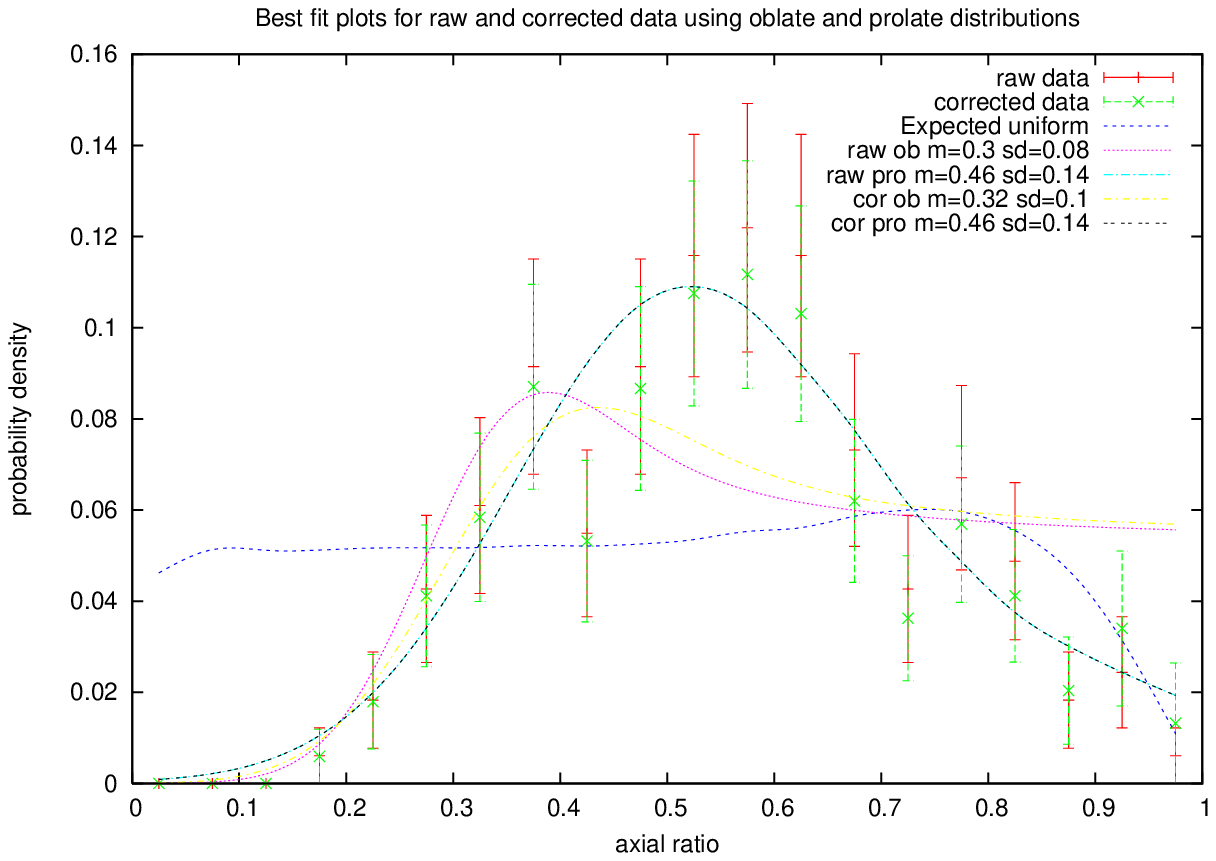}{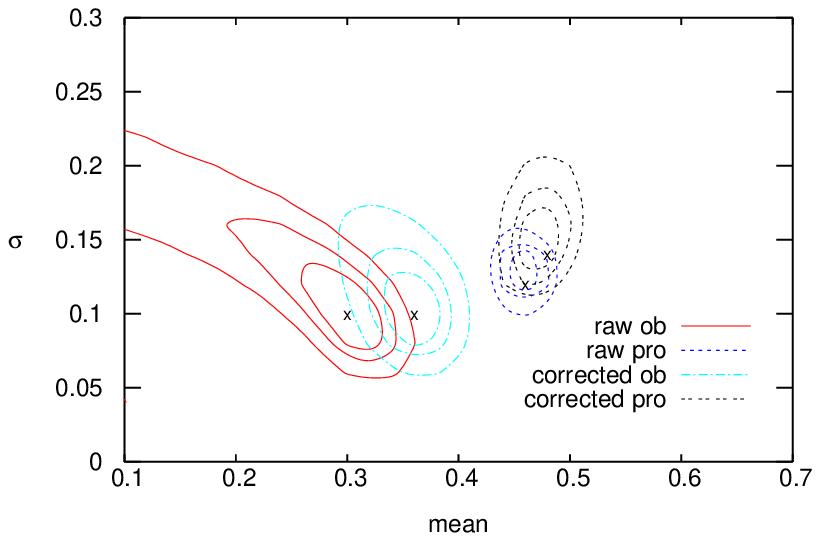}
  \caption{Group multiplicities of 20 upwards. Top plot shows best fits for the real 2PIGG data on the left and the relevant parameter contour plots on the right, whilst the bottom plot is a comparison to the mock 2PIGG data.}
  \label{fitschi20+}
\end{figure*}

The main results can be found in tables \ref{tableraw} and \ref{tablecorrect}, and figures \ref{fitschi5+}, \ref{fitschi510}, \ref{fitschi1020} and \ref{fitschi20+} for the raw and the corrected data respectively. The tables present parameter values for the best fit $\chi^2$ prolate and oblate distributions obtained, whilst the figures display the raw and corrected data overplotted with these best fit distributions along with their associated error ellipses.

\begin{deluxetable}{l|ccccc}
\tablecaption{Best fit parameters for raw group data}
\tabletypesize{\tiny}
\tablewidth{0pt}
\tablehead{
\colhead{Multiplicity} &
\colhead{$\bar{\beta}$} &
\colhead{$\sigma$} &
\colhead{$\chi^2$} &
\colhead{Reduced $\chi^2$}}
\startdata
All groups (true oblate) & $0.16$ & $0.06$ & $1719.41$ & $95.52$ \\
All groups (true prolate) & $0.36$ & $0.14$ & $74.62$ & $4.15$\\
All groups (mock oblate) & $0.14$ & $0.08$ & $1685.50$ & $93.64$ \\
All groups (mock prolate) & $0.36$ & $0.16$ & $75.96$ & $4.22$\\ \hline 
5-9 (true oblate) & $0.12$ & $0.06$ & $1836.35$ & $102.02$ \\
5-9 (true prolate) & $0.34$ & $0.14$ & $66.71$ & $3.71$\\
5-9 (mock oblate) & $0.1$ & $0.06$ & $3028.97$ & $168.28$ \\
5-9 (mock prolate) & $0.3$ & $0.14$ & $79.57$ & $4.42$\\ \hline 
10-19 (true oblate) & $0.24$ & $0.04$ & $276.33$ & $17.27$ \\
10-19 (true prolate) & $0.42$ & $0.12$ & $33.99$ & $2.12$\\
10-19 (mock oblate) & $0.26$ & $0.1$ & $194.30$ & $12.14$ \\
10-19 (mock prolate) & $0.42$ & $0.14$ & $23.60$ & $1.48$\\ \hline 
20+ (true oblate) & $0.3$ & $0.08$ & $103.66$ & $6.91$ \\
20+ (true prolate) & $0.46$ & $0.14$ & $15.97$ & $1.06 $\\
20+ (mock oblate) & $0.3$ & $0.1$ & $221.27$ & $15.81$ \\
20+ (mock prolate) & $0.46$ & $0.12$ & $23.32$ & $1.67 $\\
\enddata
\label{tableraw}
\end{deluxetable}

\begin{deluxetable}{l|ccccc}
\tablecaption{Best fit parameters for convolution corrected group data}
\tabletypesize{\tiny}
\tablewidth{0pt}
\tablehead{
\colhead{Multiplicity} &
\colhead{$\bar{\beta}$} &
\colhead{$\sigma$} &
\colhead{$\chi^2$} &
\colhead{Reduced $\chi^2$}}
\startdata
All groups (true oblate) & $0.22$ & $0.1$ & $28.47$ & $1.58$\\
All groups (true prolate) & $0.44$ & $0.18$ & $110.27$ & $6.13$\\
All groups (mock oblate) & $0.22$ & $0.14$ & $50.71$ & $2.82$\\
All groups (mock prolate) & $0.42$ & $0.18$ & $60.49$ & $3.36$\\ \hline 
5-9 (true oblate) & $0.2$ & $0.1$ & $14.47$ & $0.80$\\
5-9 (true prolate) & $0.42$ & $0.2$ & $110.07$ & $6.12$\\
5-9 (mock oblate) & $0.16$ & $0.12$ & $38.09$ & $2.12$\\
5-9 (mock prolate) & $0.4$ & $0.2$ & $65.54$ & $3.64$\\ \hline 
10-19 (true oblate) & $0.3$ & $0.1$ & $45.57$ & $2.85$\\
10-19 (true prolate) & $0.44$ & $0.14$ & $42.79$ & $2.67$\\
10-19 (mock oblate) & $0.28$ & $0.1$ & $27.00$ & $1.68$\\
10-19 (mock prolate) & $0.44$ & $0.16$ & $33.96$ & $2.12$\\ \hline 
20+ (true oblate) & $0.32$ & $0.1$ & $38.31$ & $2.55$\\
20+ (true prolate) & $0.46$ & $0.14$ & $12.74$ & $0.85$\\
20+ (mock oblate) & $0.36$ & $0.1$ & $19.03$ & $1.36$\\
20+ (mock prolate) & $0.48$ & $0.14$ & $9.91$ & $0.71$\\
\enddata
\label{tablecorrect}
\end{deluxetable}

Firstly, we will consider the raw data with a multiplicity cut-off of 20 (see figure \ref{fitschi20+}), since for the multiplicity range described it will be the most reliable raw result and immediately comparable to other research. Assuming Gaussian PDF for either oblate or prolate ellipsoids, $\chi^2$ tests were undertaken using the numerical integration capacity of Maple to an accuracy of 0.02 for both the mean (0.1 to 0.7) and the standard deviation (0.06 to 0.3). The minimum $\chi^2$ solution found the groups to be prolate to a high degree of confidence (the minimal prolate $\chi^2$ solution returned 15.97 compared to 103.66 for oblate), with a Gaussian distribution of mean $\bar{\beta}=0.46$ and standard deviation $\sigma_{\beta}=0.14$. These values agree extremely well with those recently presented by \citet{plionis06} and \citet{paz06}. For this data cut we have 15 degrees of freedom giving a reduced $\chi^2$ of 1.06, which is within $1\sigma$ expectations.

The parameter gradient for a prolate ellipsoid distribution is very steep for the raw data of groups with multiplicity between 5 and 9. The results strongly imply the distribution being narrow ($0.12<\sigma_{\beta}<0.15$), and restricted between $\bar{\beta}$ values of 0.44 to 0.48. More significantly the distribution is extremely far removed from any possible oblate distribution. This is intuitively obvious from the lack a plateau at the near circular extreme of the binned axial ratio PDF This is an effect that will always be present in oblate ellipsoids due to the observed axial ratio always having to be greater or equal to the intrinsic axial ratio (true also for prolate ellipsoids), and because an oblate ellipsoid will possess a greater observed axial ratio ($q$) when compared to a prolate ellipsoid sharing the same intrinsic axial ratio ($\beta$) inclined from our line-of-sight by the same degree. The fact the measured binned PDF falls to almost zero for circular value axial ratios strongly precludes an oblate contribution except at insignificant levels. It is due to this effect of the observed axial ratio, at a minimum, being equal to the intrinsic value that $\bar{\beta}$ will always be smaller than the peak in any observed PDF, hence in this case $\bar{\beta}$ is smaller than the obvious peak in the binned PDF at 0.6.

The other raw measurements for different multiplicity groups are presented in the corresponding graphs and tables, however such is the inherent biasing in these lower multiplicity data that little can be gleaned from their analysis alone. Instead, our last use of the raw data is to determine the consistency between the real universe and the $\Lambda$CDM dependent mock universe. It is completely justifiable to make a comparison between the raw data and their multiplicity associated mock data, since the shape biasing (being a sampling artifact) will be the same. By directly comparing the minimal $\chi^2$ fits and considering the cross catalog KS tests we will have an insight into the reliability of both the grouping algorithm and the cosmological model used. When considering table \ref{tableraw} and table \ref{kstruemock} we can see remarkably good agreement between the real and mock data \citep[c.f.][]{paz06}. Considering the fits, we find near perfect agreement for $\bar{\beta}$ between the real and mock data, the major disagreements being for multiplicities of 5-9 (for prolate distributions the real data has $\bar{\beta}=0.34$ compared to $\bar{\beta}=0.30$ for the mock data) and for 10-19 (for oblate distributions the real data has $\sigma=0.04$ compared to $\sigma=0.1$ for the mock data). We find for the most part extremely similar $\chi^2$ between the two data sets, another indication that they are truly describing the same underlying distributions. Furthermore, when looking at the KS test results we find that all direct comparisons can be drawn from the same underlying distribution to within $1\sigma$ expectations. The self similarities between the data for different multiplicities are not quite as good (tables \ref{kstrue} and \ref{ksmock}), however, the real and mock data both reject shared self similar distributions to a high degree of significance. As would be expected from the fitting results, the most similar distributions are those for multiplicities of 20+ and 10-20 for both the real and mock data ($2.9\%$ and $0.14\%$ respectively). It can be seen from the $\chi^2$ contour data that these multiplicities only agree upon parameters when considering the $2\sigma$ error contours, which is consistent with this KS test statistic.

\begin{deluxetable*}{l|cccc}
\tablecaption{KS test comparisons between true and mock data for different multiplicities}
\tabletypesize{\tiny}
\tablewidth{0pt}
\tablehead{
\colhead{} &
\colhead{Real raw All} &
\colhead{Real raw 5-9} &
\colhead{Real raw 10-19} &
\colhead{Real raw 20+}}
\startdata
Mock raw All & $68.63\%$ & $1.16x10^{-3}\%$ & $6.27\times10^{-3}\%$ & $1.55\times10^{-5}\%$\\
Mock raw 5-9 & $4.59x10^{-4}\%$ & $27.11\%$ & $5.72\times10^{-13}\%$ & $5.04\times10^{-13}\%$\\
Mock raw 10-19 & $1.61\times10^{-8}\%$ & $5.14x10^{-10}\%$ & $22.94\%$ & $0.06\%$\\
Mock raw 20+ & $3.43\times10^{-13}\%$ & $4.02\times10^{-21}\%$ & $0.07\%$ & $67.49\%$\\
\enddata
\label{kstruemock}
\end{deluxetable*}

\begin{deluxetable*}{l|cccc}
\tablecaption{KS test comparisons between self similar true data for different multiplicities}
\tabletypesize{\tiny}
\tablewidth{0pt}
\tablehead{
\colhead{} &
\colhead{Real raw All} &
\colhead{Real raw 5-9} &
\colhead{Real raw 10-19} &
\colhead{Real raw 20+}}
\startdata
Real raw All & 1 & N/A & N/A & N/A\\
Real raw 5-9 & $0.04\%$ & 1 & N/A & N/A\\
Real raw 10-19 & $1.11\times10^{-3}\%$ & $5.35\times10^{-10}$ & 1 & N/A\\
Real raw 20+ & $1.83\times10^{-6}\%$ & $8.27\times10^{-12}\%$ & $2.90\%$ & 1\\
\enddata
\label{kstrue}
\end{deluxetable*}

\begin{deluxetable*}{l|cccc}
\tablecaption{KS test comparisons between self similar mock data for different multiplicities}
\tabletypesize{\tiny}
\tablewidth{0pt}
\tablehead{
\colhead{} &
\colhead{Mock raw All} &
\colhead{Mock raw 5-9} &
\colhead{Mock raw 19-20} &
\colhead{Mock raw 20+}}
\startdata
Mock raw All & 1 & N/A & N/A & N/A\\
Mock raw 5-9 & $2.45\times10^{-5}\%$ & 1 & N/A & N/A\\
Mock raw 10-19 & $1.97\times10^{-3}\%$ & $1.06\times10^{-12}$ & 1 & N/A\\
Mock raw 20+ & $1.44\times10^{-11}\%$ & $4.056\times10^{-23}\%$ & $0.14\%$ & 1\\
\enddata
\label{ksmock}
\end{deluxetable*}

We now consider the convolution corrected data with weighted histogram bins. The results for the real and mock forms of the 2PIGG catalog can be seen in figures \ref{fitschi5+}, \ref{fitschi510}, \ref{fitschi1020} and \ref{fitschi20+}, and in table \ref{tablecorrect}. Significantly, we now find that the distributions are not necessarily prolate. The complete data and the multiplicity cut for 5-9 is better fit by an oblate distribution, the cut-off for 10-19 is equally well fit by either, and only the 20+ cut is better fit by a prolate distribution and not with a definite rejection of an oblate distribution. If oblate distributions are accepted for the corrected data then we see a strong divide between the low multiplicity cut (5-9) and the other two larger multiplicity cuts. The former has a distribution mean $\bar{\beta}=0.2$ whilst the latter two have a mean $\bar{\beta}\sim0.3$: in all cases $\sigma_{\beta}=0.1$. If we still assume prolate distributions, despite evidence of strong rejection in our 5-9 multiplicity cut, then we still see a small variation in distribution mean: $\bar{\beta}=0.42\pm0.01$ for 5-9, $\bar{\beta}=0.44\pm0.01$ for 10-19 and $\bar{\beta}=0.46\pm0.02$ for 20+. The latter two have $\sigma_{\beta}=0.14$, whilst the 5-9 cut has $\sigma_{\beta}=0.2$. Within the $1\sigma$ error contours the two higher multiplicity cuts share the same distribution, $\bar{\beta}\sim0.44$ and $\sigma_{\beta}=0.14$, whilst the 5-9 cut is strongly indicated to be a truly different distribution, lying outside the $3\sigma$ error contours. This finding is consistant with the previous KS tests which showed the two high multiplicity cuts to be the most similar; once the distributions are corrected they become identical. Evidence of oblate shapes has been presented before in literature: \citet{fasano93} and \citet{orlov01} being notable examples, though with a smaller sample size and different multiplicity limits respectively. These authors also found a non-rejectable oblate distribution, although \cite{fasano93} favoured the prolate fit.
\cite{orlov01} found an axial ratio mean of 3:1 for their sample of small galaxy groups (multiplicities of 3-8), a figure that can be considered consistent with our data ($\bar{\beta}=0.2$ for group multiplicities 5-9) given that the statistics are significantly poorer (2703 galaxies in 485 groups). Prolate fits have been found to be the best fit for large cluster samples \citep{cooray00} which would be consistant with our 20+ multiplicity cut (these systems being the most comparable to cluster environments).

Regarding comparisons to the mock catalog we once again see consistent fitting between the real and mock data, accounting for the corresponding error ellipses the fitting between the real and mock data is actually better than for the raw data. However, the mock catalogs do not find such similar distributions for the 10-19 and 20+ cuts, either for oblate ($\bar{\beta}=0.28$, $\bar{\beta}=0.36$ respectively) or prolate ($\bar{\beta}=0.44$, $\bar{\beta}=0.48$ respectively) solutions. It should be noted that a KS test cannot be done in this situation since it is only applicable to data with unweighted histogram frequency points.

Perhaps the most interesting aspect of this analysis is the tentative evidence it shows for generally more spherical groups for larger multiplicities, not just as a numerical artifact. This would be consistant with large multiplicity groups being more virialised and dynamically more evolved, consistant with hierarchical formation of small groups collapsing along filaments and larger groups forming at nodes. The features that account for the high ellipticity sub sample in low multiplicity groups are a low $\bar{\beta}$ for an oblate solution, and a large $\sigma$ for a prolate solution. Either way there does appear to be a low axial ratio population that must be accounted for, and that is not present in the higher multiplicity data. It is worth remembering that at low magnitude cuts (corresponding to higher multiplicities) the Local Group demonstrates oblate characteristics, so this result is not necessarily suprising. This result is even more reliable occurring as it does in the region of the PDF that is least corrected in convolution. In fact in this region highly elliptical results for low multiplicity groups are reduced in weighting, so the fact the signal is still so strong indicates a real population difference. This trend would appear to be in the opposite sense to what is expected in simulations (see \citep{allgood06} and references therein), but it should be noted that these results are much closer to agreement after correction.

\section{Conclusion}

We found good agreement between our results for raw data with a simple 20+ multiplicity cut and findings by other authors, indicating a strongly prolate distribution with a mean $\bar{\beta}=0.46$ and $\sigma_{\beta}=0.14$. However, when fits were applied to convolution corrected distributions which allow for the error introduced by finite sampling we found that for the higher multiplicity cuts both oblate and prolate distributions could produce reasonable fits. More significantly, evidence suggests that these large multiplicity populations share the same underlying distribution (if oblate: $\bar{\beta}\sim0.3$ $\sigma_{\beta}=0.1$, if prolate $\bar{\beta}\sim0.44$ $\sigma_{\beta}=0.14$). There is evidence in the data that low multiplicity groups have a highly elliptical sub population that is missing in other multiplicity cuts: for 5-9 multiplicity cuts oblate distributions produce a better quality of fit ($\bar{\beta}=0.2$, $\sigma_{\beta}=0.1$), this is despite the effect of the convolution correction being to reduce this signal. This is seen as evidence of high multiplicity groups being more spherical, and is consistant with them being located in nodes, or at least in a less filamentary structure than the extremely low multiplicity groups. Large multiplicity groups are better fit by a prolate distribution in the corrected data, but oblate fits are not as strongly rejected as in earlier work which did not account for the sampling bias. The effect of interlopers was extensively considered, and the results of this work imply all measurements obtained should be considered upper limits for both the under lying axial ratios (i.e. the true non-interloper distribution will be more elliptical than that measured) and the standard deviations of the populations.

KS tests were utilised where appropriate and indicate that the real and mock catalogs are in very good agreement over all ranges. For all multiplicity comparisons the real and mock data share the same underlying distribution to within $1\sigma$ expectations, and the most self similar populations are for multiplicities 10-19 and 20+, which appear consistent with being part of the same overall population once the corrections have been applied.

Further exploration of projected group shapes will be presented in a future paper which will consider the differences we find when colour cuts and different grouping algorithms are used.

\section*{Acknowledgements}
AR acknowledges funding through the UK Particle Physics and Astrophysics
Research Council (PPARC). We would like to thank the referee for their helpful suggestions, particularly the suggestion to discuss the effect of interlopers in more detail.

\section*{Appendix}

For a prolate or oblate ellipsoid we know that
\begin{equation}
(ux)^2+(uy)^2+z^2=a^2
\end{equation}
where $u>1$ for a prolate ellipsoid, and $u<1$ for an oblate ellipsoid. So in z-x plane
\begin{equation}
(ux_0)^2+z_0^2=a^2
\end{equation}

\begin{equation}
z_0^2=a^2-(ux_0)^2
\end{equation}

\begin{equation}
z_0^2.2\frac{\delta z_0}{\delta x_0}=-2u^2x_0
\end{equation}
>From standard trigonometry
\begin{mathletters}
\begin{equation}
\tan\left(\frac{\pi}{2}-\theta\right)=\cot(\theta)
\end{equation}
so the gradient with respect to the observer is defined by
\begin{equation}
\frac{\delta z_0}{\delta x_0}=-\frac{u^2 x_0}{z_0}=\cot\theta
\end{equation}
and
\begin{equation}
\frac{x_0^2}{z_0^2}=\frac{a^2}{u^2z_0^2}-\frac{1}{u^2}=\frac{1}{u^2}\left(\frac{a^2}{z_0^2}-1\right)
\end{equation}
giving us
\begin{equation}
\cot^2\theta=\frac{u^4x_0^2}{z_0^2}=u^2\left(\frac{a^2}{z_0^2}-1\right)
\label{cot}
\end{equation}
\end{mathletters}
>From figure \ref{ellipse} we can see
\begin{equation}
\frac{A}{C}=\sin\theta
\end{equation}

\begin{equation}
C=\frac{A}{\sin\theta}=z_0-x_0.m
\end{equation}
where m is the gradient, and will necessarily be the opposite sign to $x_0$. This gives the z value for the z-axis/tangent intercept.
\begin{equation}
C=z_0-x_0\frac{-u^2x_0}{z_0}=z_0+\frac{u^2x_0^2}{z_0}=\frac{a^2}{z_0}
\end{equation}
Using (\ref{cot}) we get
\begin{equation}
A^2=a^2\left(\frac{a^2}{z_0^2}\right)\sin^2\theta=a^2\left(\frac{\cos^2\theta}{u^2}+\sin^2\theta\right)
\end{equation}
We define the apparent axial ratio ($q$) to be less than 1 for both prolate and oblate ellipsoids. Thus for oblate ellipsoids $q=\frac{uA}{a}$ and for prolate ellipsoids $q=\frac{a}{uA}$. So
\begin{equation}
u^2\sin^2\theta+\cos^2\theta=q^2 (oblate) = \frac{1}{q^2} (prolate)
\end{equation}
To have $q$ between $q$ and $q+\delta q$, the symmetry axis has to lie at an angle between $\theta$ and $\theta+\delta\theta$.
\begin{equation}
\delta\theta=\frac{\delta q}{|\delta q/\delta\theta|}
\end{equation}
Mathematically this means if q changes rapidly with respect to $\theta$ then $\delta\theta$ is small. A simple example is to imagine a needle (an almost completely prolate object): when we view this end on and alter the angle, $\delta q/\delta\theta$ will be huge; our observed axial ratio $q$ has changed from 1 (a perfect circle) to near 0 (by this definition almost a line) over an infinitesimally small angle. So the range of angles over which we might say the axial ratio is similar $(\delta\theta)$ is very small. However, when we consider the same needle side on, and rotate as before, $\delta q/\delta\theta$ will be very small; $q$ is changing very gradually. So for a prolate object viewed with the symmetry axis at an orthogonal angle, the range of angles possessing similar axial ratios is much larger than when viewed with the symmetry axis parallel to our line-of-sight.

For a given $u$, and assuming a random orientation of symmetry axes for all the ellipsoids, the relative probability of observing the ellipsoid between two axial ratios (say $q_1$ and $q_2$) is given by the relative magnitude of the cosine between angles $\theta_1$ and $\theta_2$ that correspond to these observed ellipsoids. This is exactly the same factor that has to be considered when trying to determine what fraction of the angular area of the celestial sphere is contained within different declinations. As an example, if $q$ varies the same amount between $0^\circ$ and $10^\circ$ as it does between $80^\circ$ and $90^\circ$, the difference in the solid angle of sky subtended will be $\frac{\cos 80^\circ}{1-\cos 10^\circ}$=11.43; thus it is 11.43 times more likely we will see the ellipsoid in the larger $\theta$ range. Taking this to infinitesimal differences we have
\begin{equation}
\left|\cos(\theta+\delta\theta)-\cos\theta\right|=\left|\cos\theta\cos\delta\theta-\sin\theta\sin\delta\theta -\cos\theta\right|=\sin\theta\delta\theta
\end{equation}
So of the ellipsoids under consideration, a fraction $\sin\theta\delta\theta$ will have their symmetry axes directed at an angle $\theta$ to the line-of-sight. Now we can write
\begin{equation}
\sin\theta\delta\theta=\frac{\sin\theta\delta q}{|\delta q/\delta\theta|}
\end{equation}
Galaxies with $u$ in the range $(u,u+\delta u)$ contribute to the observed $f(q)\delta q$ galaxies with axial ratios in the range $(q,q+\delta q)$ accordingly
\begin{equation}
f(q)\delta q=\frac{n(u)\delta u\sin\theta\delta q}{|\delta q/\delta\theta|}
\end{equation}
So the probability density function (PDF) of observed axial ratios ($q$) is given by
\begin{equation}
f(q)=\int n(u)\left(\frac{\sin\theta}{|\delta q/\delta\theta|}\right)du
\label{fq}
\end{equation}
For an oblate ellipsoid (from Eq. \ref{trigdef})
\begin{equation}
u^2(1-\cos^2\theta)+\cos^2\theta=q^2
\end{equation}
$\therefore$
\begin{equation}
\cos^2\theta=\frac{q^2-u^2}{1-u^2}
\end{equation}
Similarly
\begin{equation}
u^2\sin^2\theta+1-\sin^2\theta=q^2
\end{equation}
\begin{equation}
sin^2\theta=\frac{1-q^2}{1-u^2}
\label{sine}
\end{equation}
It follows that
\begin{equation}
\frac{\delta q}{\delta\theta}=-\frac{(q^2-u^2)^{0.5}}{(1-u^2)^{0.5}}\frac{(1-q^2)^{0.5}}{(1-u^2)^{0.5}}\frac{1-u^2}{q}
\end{equation}
$\therefore$
\begin{equation}
\left|\frac{\delta q}{\delta\theta}\right|=\frac{1}{q}[(1-q^2)(q^2-u^2)]^{0.5}
\label{diff}
\end{equation}
Using equations (\ref{fq}), (\ref{sine}) and (\ref{diff}) we have
\begin{equation}
f(q)=q\int^q_0\frac{n(u)du}{[(1-u^2)(q^2-u^2)]^{0.5}}
\end{equation}
We now define the intrinsic axial ratio to be a value $\beta$ always less than 1, for an oblate ellipsoid $\beta=u$. $f(q)$ alone give us a PDF of the observed axial ratios, however our data will have to be binned for analysis. So for a given value of $\beta$ we integrate over a range of observed axial ratios (say $q_1$ to $q_2$) to construct a number density function for observed axial ratios $N(\beta,q_1,q_2)$ where
\begin{equation}
N(\beta,q_1,q_2)=\frac{1}{\sqrt{1-\beta^2}}\int^{q_2}_{q_1}\frac{q dq}{\sqrt{q^2-\beta^2}}
\label{intref}
\end{equation}
Considering the integral alone we use the substitution $q=\beta\cos\theta$ (where $\delta q=-\beta\sin\theta\delta\theta$), giving us
\begin{equation}
\int^{q_2}_{q_1}-\frac{\beta^2\cos\theta\sin\theta d\theta}{\sqrt{\beta^2\cos^2\theta-\beta^2}}=\int^{q_2}_{q_1}-\frac{\beta^2\cos\theta\sin\theta d\theta}{i\beta\sin\theta}=\int^{q_2}_{q_1}i\beta\cos\theta d\theta=\left[i\beta\sin\theta+C\right]^{q_2}_{q_1}
\label{intref2}
\end{equation}
Rearranging the original substitution and replacing $\sin\theta$ accordingly, the integral becomes
\begin{equation}
\left[\sqrt{q^2-\beta^2}\right]^{q_2}_{q_1}
\end{equation}
so for an oblate ellipsoid with a given $\beta$
\begin{equation}
N(\beta,q_1,q_2)=\frac{1}{\sqrt{1-\beta^2}}\left[\sqrt{(q_2^2-\beta^2)}-\sqrt{(q_1^2-\beta^2)}\right]
\end{equation}
For a prolate ellipsoid $\beta=\frac{1}{u}$ so the integral in (\ref{intref}) is multiplied by a factor $\frac{\beta^2}{q^3}$, thus \ref{intref2} becomes
\begin{equation}
\int^{q_2}_{q_1}-\frac{\beta^3\sin\theta\delta\theta}{\beta^2\cos^2\theta(\sqrt{\beta^2\cos^2\theta-\beta^2})}=\int^{q_2}_{q_1}-\frac{\delta\theta}{i\cos^2\theta}=\left[i\frac{\sin\theta}{\cos\theta}+C\right]^{q_2}_{q_1}
\end{equation}
so for a prolate ellipsoid with a given $\beta$
\begin{equation}
N(\beta,q_1,q_2)=\frac{1}{\sqrt{1-\beta^2}}\left[\frac{\sqrt{q_2^2-\beta^2}}{q_2}-\frac{\sqrt{q_1^2-\beta^2}}{q_1}\right]
\end{equation}
the only difference is a factor of the apparent axial ratio being considered, (this makes sense because you would expect a prolate object to have a larger number density of small observed axial ratios). Putting this all together, for a function $\tilde{n}(\beta)$ using oblate ellipsoids we find
\begin{equation}
N(q_1,q_2)=\int^{q_2}_0\tilde{n}(\beta)\frac{1}{\sqrt{1-\beta^2}}\left[\sqrt{(q_2^2-\beta^2)}-\sqrt{(q_1^2-\beta^2)}\right]d \beta
\end{equation}
and for prolate ellipsoids
\begin{equation}
N(q_1,q_2)=\int^{q_2}_0\tilde{n}(\beta)\frac{1}{\sqrt{1-\beta^2}}\left[\frac{\sqrt{(q_2^2-\beta^2)}}{q_2}-\frac{\sqrt{(q_1^2-\beta^2)}}{q_1}\right]d \beta
\end{equation}

\clearpage


%% If you are not including electonic art with your submission, you may
%% mark up your captions using the \figcaption command. See the 
%% No more than seven \figcaption commands are allowed per page, 
%% so if you have more than seven captions, insert a \clearpage 

%% Tables should be submitted one per page, so put a \clearpage before

%% Two options are available to the author for producing tables:  the
%% deluxetable environment provided by the AASTeX package or the LaTeX
%% table environment.  Use of deluxetable is preferred.
%%

%% Three table samples follow, two marked up in the deluxetable environment,
%% one marked up as a LaTeX table.

\end{document}